\documentclass[aps,twocolumn,prb,showpacs,floatfix]{revtex4}

\usepackage{epsfig,amsmath,amssymb}
\usepackage{graphics,color} 
\usepackage{subfigure}

\usepackage{multirow}

\newcommand{\plusontimes}{\rlap{$+$}$\times$}

\begin{document}

\title{Phase diagram of a frustrated Heisenberg antiferromagnet on the honeycomb lattice:
    \\the $J_{1}$--$J_{2}$--$J_{3}$ model}

\author{P. H. Y. Li and R. F. Bishop}

\affiliation{School of Physics and Astronomy, Schuster Building, The University of Manchester, 
Manchester M13 9PL, United Kingdom}

\author{D. J. J. Farnell}

\affiliation{Division of Mathematics, Faculty of Advanced Technology, 
University of Glamorgan, Pontypridd CF37 1DL, Wales, United Kingdom}

\author{C.~E.~Campbell}
\affiliation{School of Physics and Astronomy, University of Minnesota, 116 Church Street SE, Minneapolis, Minnesota 55455, USA}

\date{\today}

\begin{abstract}

  We use the coupled cluster method in high orders of approximation
  to make a comprehensive study of the ground-state (GS) phase
  diagram of the spin-$\frac{1}{2}$ $J_{1}$--$J_{2}$--$J_{3}$ model
  on a two-dimensional honeycomb lattice with antiferromagnetic (AFM)
  interactions up to third-nearest neighbors.  Results are presented
  for the GS energy and the average local on-site magnetization.
  With the nearest-neighbor coupling strength $J_{1} \equiv 1$ we find
  four magnetically ordered phases in the parameter window
  $J_{2},J_{3} \in [0,1]$, namely the N\'{e}el (N), striped (S), and
  anti-N\'{e}el (aN) collinear AFM phases, plus a spiral phase.
  The aN phase appears as a stable GS phase in the classical version of the
  model only for values $J_{3}<0$.  Each of these four ordered phases
  shares a boundary with a disordered quantum paramagnetic (QP) phase,
  and at several widely separated points on the phase boundaries the 
  QP phase has an infinite susceptibility to plaquette valence-bond
  crystalline order.  We identify all of the phase boundaries with
  good precision in the parameter window studied, and we find
  three tricritical quantum critical points therein at: (a)
  $(J_{2}^{c_1},J_{3}^{c_1})=(0.51 \pm 0.01,0.69 \pm 0.01)$ between
  the N, S, and QP phases; (b) $(J_{2}^{c_2},J_{3}^{c_2})=(0.65
  \pm 0.02,0.55 \pm 0.01)$ between the S, spiral, and QP phases; and
  (c) $(J_{2}^{c_3},J_{3}^{c_3})=(0.69 \pm 0.01,0.12 \pm 0.01)$ between the
  spiral, aN, and QP phases.
  
\end{abstract}


\maketitle

\section{INTRODUCTION}

Phase transitions are common phenomena in nature and they have long been
the subject of intense theoretical interest.  Since they are driven by thermal 
and/or quantum fluctuations, an understanding of them at the microscopic level
requires a good many-body description.  Particular interest in recent years
has centered on the so-called quantum phase transitions that occur
in systems in their ground state at zero temperature ($T=0$) as some
parameters in the Hamiltonian describing them are varied.  

Spin-lattice
systems with competing interactions provide a particularly rich field in
which to study such quantum phase transitions.  The exchange interactions 
that lead to collective magnetic behavior in spin systems are clearly 
quantum-mechanical in origin.  One knows too that the interplay
between reduced dimensionality, frustration (due to either competing 
interactions or the underlying lattice geometry), and
strong quantum fluctuations generates a huge variety of new
states of condensed matter beyond the usual states of quasiclassical
long-range order (LRO).  A particularly rich place to observe such exotic ground-state 
(GS) phases is in situations where they originate as a result of quantum
fluctuations within a large set of configurations that are degenerate
at the classical level.\cite{Ramirez:2008,Balents:2010}

The search for such exotic phases that owe their
existence purely to quantum effects is nowadays one of the primary
reasons for the study of frustrated quantum spin-lattice systems.
The interplay between magnetic frustration and quantum
fluctuations provides a powerful mechanism for disturbing,
destabilizing, or even completely destroying magnetic order.  The search 
for a genuine quantum spin-liquid (QSL) phase,\cite{Ramirez:2008,Balents:2010,Anderson:1973,Anderson:1987} 
which has no magnetic order and no LRO or long-range correlations of any kind, 
has itself attracted huge theoretical interest ever since its first proposal
nearly 40 years ago by Anderson,\cite{Anderson:1973} and its subsequent
pursuit by, for example, Shastry and Sutherland\cite{shastry1}, and many others
following them.  

The GS wave function of a QSL is clearly a superposition
of a large number of different configurations.  An example could be the resonating valence-bond (RVB)
state, which is a superposition of many short-range singlet valence bonds.
The RVB state was itself first proposed by Fazekas and Anderson\cite{Fazekas:1974} as the
GS wave function for the spin-$\frac{1}{2}$ isotropic Heisenberg antiferromagnet (HAFM) on
the geometrically frustrated, two-dimensional, (2D) triangular lattice with only
nearest-neighbor (NN) interactions, all of equal strength, although it is now known
that this proposal is incorrect and that this system is magnetically ordered.

In a genuine QSL state no symmetry is broken and 
the quantum fluctuations are required to form a many-body
singlet state that contains no long-range correlations with respect to {\it any} operator,
although there may be present some form of topological order. Other exotic 
quantum paramagnetic (QP) states, which also have no magnetic LRO but which break some spatial symmetry 
with respect to short-range magnetic correlations, can also arise.  The various QP valence-bond crystalline (VBC)
solid phases fall into this category.

As we have noted, a combination of strong quantum fluctuations and strong frustration in a spin system 
provides an ideal scenario for the emergence of such novel quantum GS phases as the QSL and
other QP phases discussed above, which do not possess the magnetic LRO that typifies the
classical GS phases of the corresponding models taken in the limit $s \to \infty$ of 
the spin quantum number $s$ of the lattice spins.  We know that
quantum fluctuations tend to be largest for the smallest values of $s$, for lower
dimensionality $D$ of the lattice, and for the smallest coordination number $z$ of the lattice. Thus,
for spin-$\frac{1}{2}$ models the honeycomb lattice plays a special role since 
its coordination number ($z=3$) is the lowest possible for $D=2$.  Frustration is
easily incorporated by the inclusion of competing next-nearest-neighbor (NNN) 
and possibly also next-next-nearest-neighbor (NNNN) bonds.  For these reasons such 
spin-$\frac{1}{2}$ frustrated Heisenberg models on the honeycomb lattice have engendered 
huge theoretical interest.\cite{Rastelli:1979,Mattson:1994,Fouet:2001,Mulder:2010,
Cabra:2011,Ganesh:2011,Clark:2011,Reuther:2011,DJJF:2011_honeycomb,Albuquerque:2011,
Mosadeq:2011,Oitmaa:2011,Mezzacapo:2012,PHYLi:2012_honeycomb_J1neg,
PHYLi:2012_honeyJ1-J2,PHYLi:2012_Honeycomb_J2neg}

Interest in the honeycomb lattice has been given further impetus by the discovery of
a QSL phase in the exactly solvable Kitaev model,\cite{kitaev} in
which the spin-$\frac{1}{2}$ particles are sited on a honeycomb lattice.
Additional interest has also emanated from the recent
synthesis of graphene monolayers\cite{graphene} and other magnetic materials with a honeycomb
structure. For example, it is likely that Hubbard-like models on the honeycomb lattice may well
describe many of the physical properties of graphene.  In this context it is
particularly interesting to note the clear
evidence of Meng {\it et al.}\cite{meng} that quantum fluctuations are
strong enough to trigger an insulating QSL phase between the
non-magnetic metallic phase and the antiferromagnetic (AFM) Mott insulator for the Hubbard
model on the honeycomb lattice at moderate values of the Coulomb
repulsion $U$.  This latter Mott insulator phase corresponds 
in the limit $U \to \infty$ to the pure HAFM on the bipartite
honeycomb lattice, whose GS phase exhibits N\'{e}el LRO.  However, higher-order terms
in the $t/U$ expansion of the Hubbard model (where $t$ is the Hubbard hopping term 
strength parameter) lead to frustrating exchange
couplings in the corresponding spin-lattice limiting model (and see, 
e.g., Ref. [\onlinecite{Yang:2011_hcomb}]),
in which the HAFM with NN exchange couplings is the leading term in the large-$U$
expansion.

The unexpected result of Meng {\it et al.},\cite{meng} together with other related 
work,\cite{Yang:2011_hcomb,Vaezi:2010,Vaezi:2011} has excited much
interest in understanding the physics of frustrated quantum magnets on the honeycomb lattice. 
In particular, a growing consensus is emerging\cite{Mattson:1994,Fouet:2001,Cabra:2011,Mosadeq:2011,
Albuquerque:2011,Reuther:2011,Mezzacapo:2012} that frustrated spin-$\frac{1}{2}$ HAFMs on the honeycomb
lattice exhibit a frustration-induced QP phase.  It is interesting to note in
this context that recent experiments\cite{exp} on the
layered compound Bi$_3$Mn$_4$O$_{12}$(NO$_3$) (BMNO) at temperatures below its
Curie-Weiss temperature reveal QSL-like behavior. 
In BMNO the Mn$^{4+}$ ions are situated on the sites of (weakly-coupled) honeycomb lattices, 
although they have spin quantum number $s = \frac{3}{2}$.  The successful
substitution of the $s=\frac{3}{2}$ Mn$^{4+}$ ions in BMNO by V$^{4+}$ ions could
lead to a corresponding experimental realization of a spin-$\frac{1}{2}$ HAFM on the
honeycomb lattice.

Other realizations of quantum HAFMs which exhibit the
honeycomb structure include magnetic compounds such as
InNa$_{3}$Cu$_{2}$SbO$_{6}$\cite{Miura:2006} and
InCu$_{2/3}$V$_{1/3}$O$_{3}$.\cite{Kataev:2005}  In both of these
materials the Cu$^{2+}$ ions in the copper oxide layers form a
spin-$\frac{1}{2}$ HAFM on a (distorted) honeycomb lattice.
Other similar honeycomb materials
include the family of compounds BaM$_2$(XO$_4$)$_2$ (M=Co, Ni; X=P, As),\cite{Regnault:1990} in which
the magnetic ions M are disposed in weakly-coupled layers where they are situated on the
sites of a honeycomb lattice.  The Co ions have spins $s = \frac{1}{2}$ and the Ni ions have spins
$s = 1$. Recent calculations\cite{Tsirlin:2010} of the material $\beta$-Cu$_2$V$_2$O$_7$
have demonstrated that its properties can also be described in terms of
a spin-$\frac{1}{2}$ model on an (anisotropic) honeycomb lattice.

Finally, we note that the very recent prospect of being able to realize spin-lattice models
with ultracold atoms trapped in optical lattices\cite{Struck:2011} is
likely to make even more data available about the quantum phase transitions
in the models as the exciting possibility opens up in such trapped-atom
experiments to tune the strengths of the competing magnetic bonds,
and hence to drive the system from one phase to another.

Recently, we have made a series of studies of the frustrated
spin-$\frac{1}{2}$ $J_{1}$--$J_{2}$--$J_{3}$ model on the honeycomb lattice using the
coupled cluster method (CCM) complemented in some cases with the
Lanczos exact diagonalization (ED) of small lattices.  We have studied various regimes
in the full $(J_{1},J_{2},J_{3})$ parameter space of NN ($J_1$) bonds, NNN ($J_2$) bonds,
and NNNN ($J_3$) bonds.  These include (a) the AFM model (i.e., with $J_{1}>0$) in the special 
case where the NNN and NNNN bonds are also AFM and have equal strength 
($J_{3}=J_{2} \equiv \kappa J_{1} > 0$);\cite{DJJF:2011_honeycomb} (b) the ferromagnetic (FM) model
(i.e., with $J_{1}<0$) with frustrating NNN and NNNN bonds, again in the special case
where they are both AFM and have equal strength ($J_{3}=J_{2}>0$);\cite{PHYLi:2012_honeycomb_J1neg} 
(c) the AFM model (i.e., with $J_{1}>0$) in the special case where the NNN and NNNN bonds
are both FM  and have equal strength ($J_{3} = J_{2} < 0$);\cite{PHYLi:2012_Honeycomb_J2neg} 
and (d) the AFM model (i.e., with $J_{1}>0$) in the special case where we have frustrating 
NNN bonds only (i.e., with $J_{2} \equiv x J_{1}>0$; $J_{3}=0$).\cite{PHYLi:2012_honeyJ1-J2}
The aim of the present work is to extend the investigation  of the phase diagram of the full
spin-$\frac{1}{2}$ $J_{1}$--$J_{2}$--$J_{3}$ model on the honeycomb lattice in the case where all of
the NN, NNN, and NNNN bonds are AFM, but no further restriction is made except that we
limit the parameter space window to $J_{2}/J_{1},J_{3}/J_{1} \in [0,1]$.

We briefly outline the structure of the rest of the paper.  The
model itself is first described in Sec.~\ref{model_section}, before 
briefly outlining the CCM formalism that we employ as our main 
calculational tool in Sec.~\ref{CCM}.  To aid the reader we first give an overview
of our main results in Sec.~\ref{preview}, focussing on the phase
diagram for the model, before we give a detailed presentation and
discussion of our results in Sec.~\ref{results}.  Finally, we conclude 
in Sec.~\ref{summary} with a summary and comparison of our results with the work
of others.

\section{THE HONEYCOMB MODEL}
\label{model_section}
\begin{figure*}[!tb]
\mbox{
\subfigure[]{\scalebox{0.3}{\includegraphics{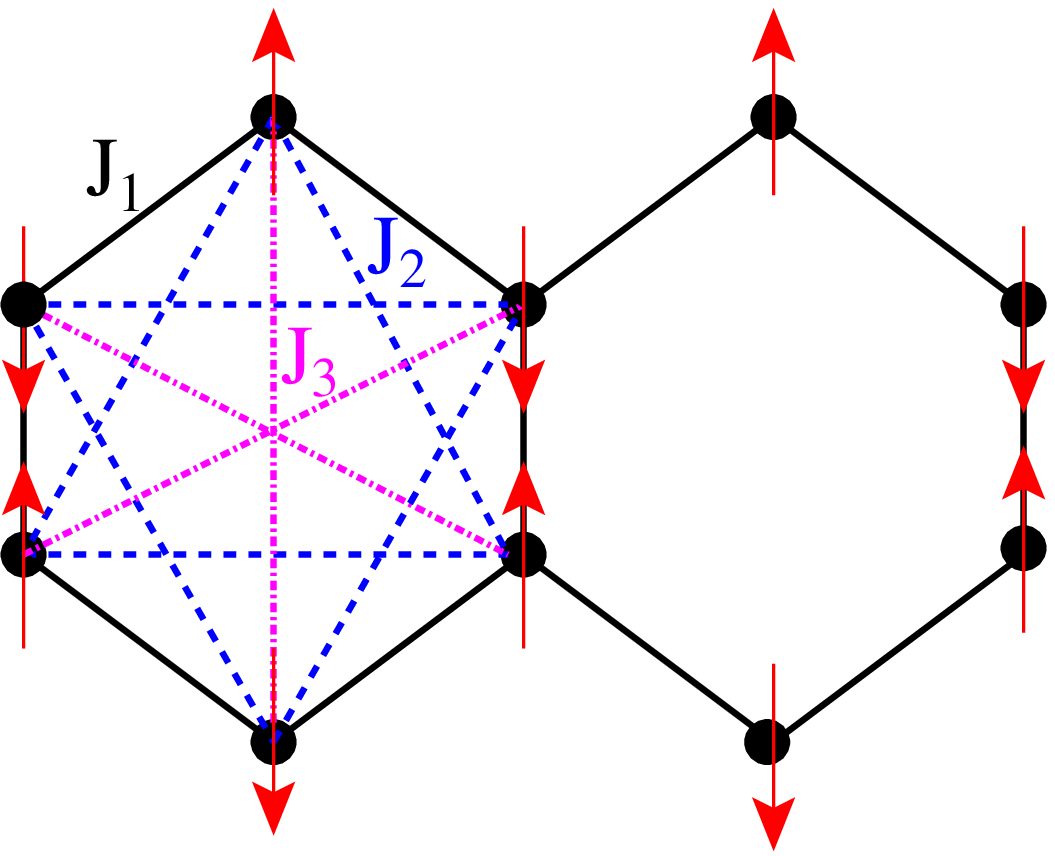}}}
\qquad \subfigure[]{\scalebox{0.3}{\includegraphics{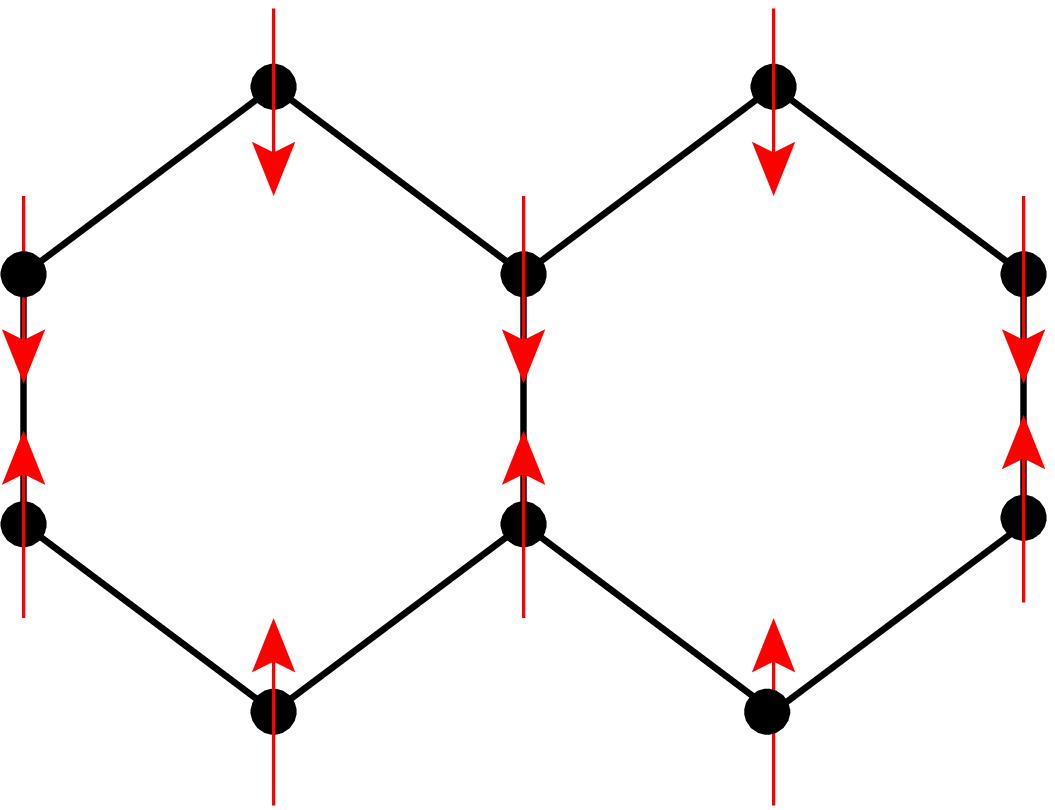}}}
\qquad \subfigure[]{\scalebox{0.3}{\includegraphics{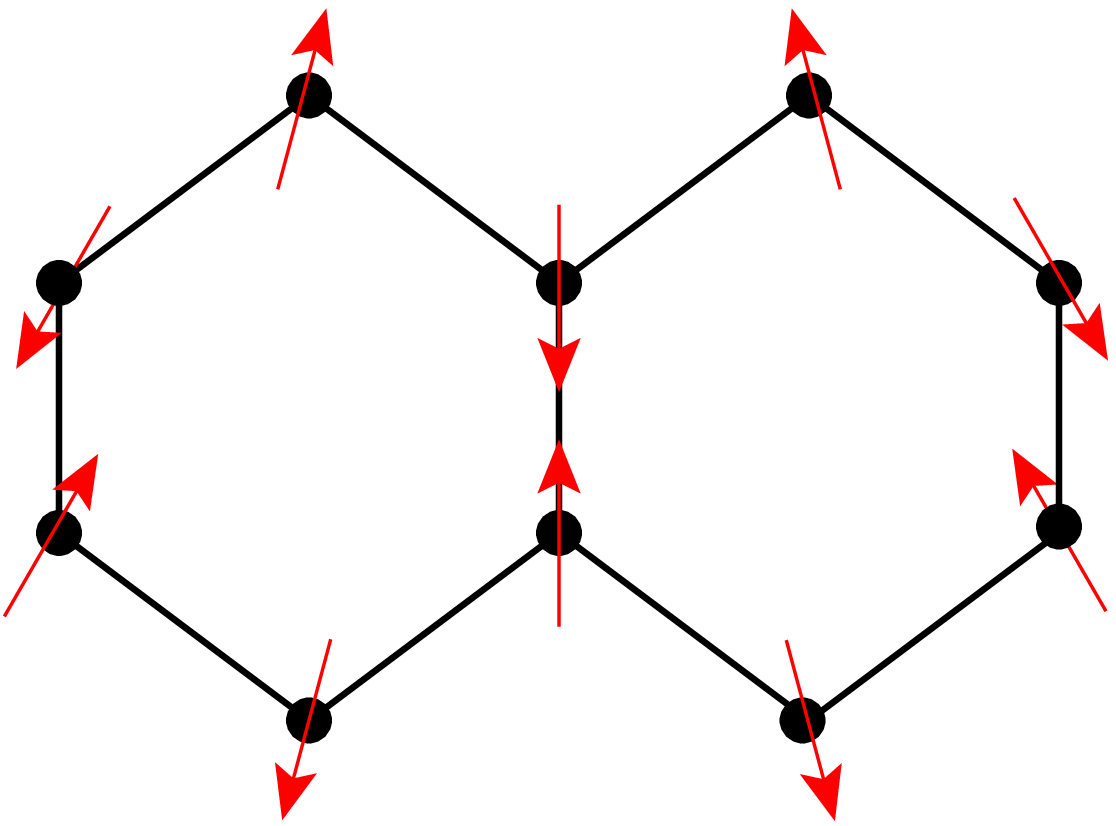}}}
\qquad \subfigure[]{\scalebox{0.3}{\includegraphics{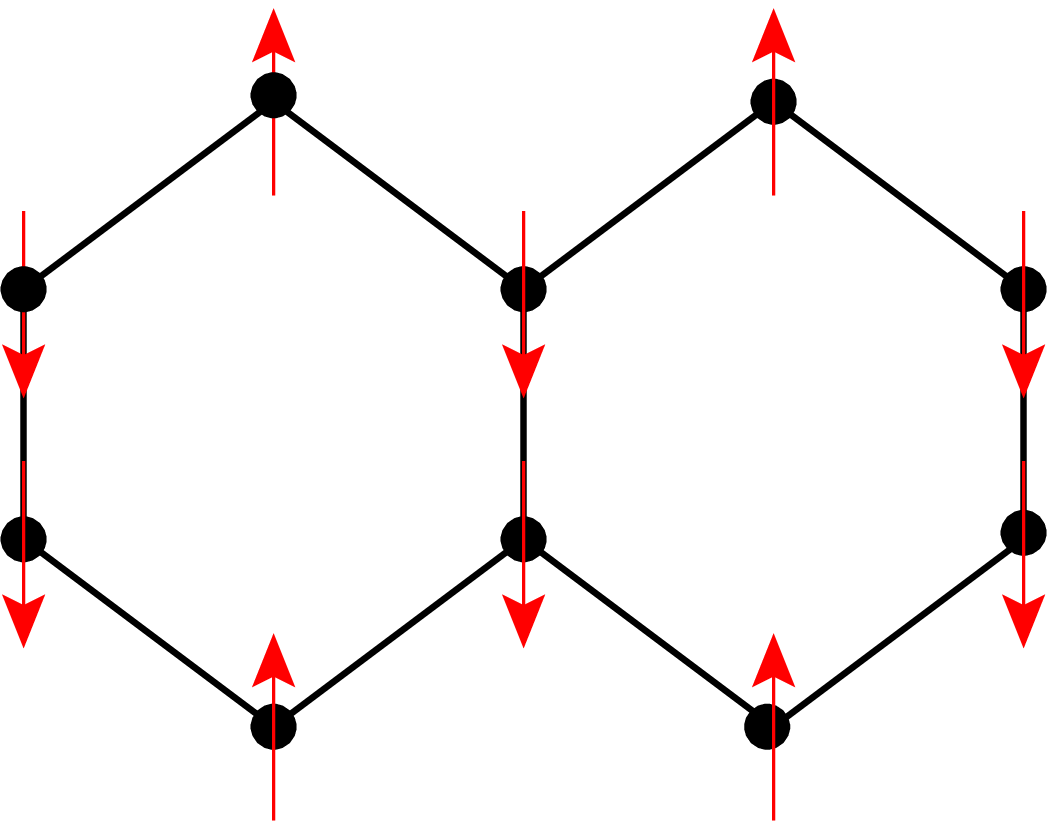}}}
}
\caption{(Color online) The $J_{1}$-$J_{2}$-$J_{3}$ honeycomb model
  with $J_{1} \equiv 1; J_{2}>0; J_{3}>0$, showing
  the (a) N\'{e}el (N), (b) striped (S), (c) spiral and (d) anti-N\'{e}el (aN) states.
  The spins on lattice sites \textbullet \hspace{0.01cm} are represented by the (red) arrows.}
\label{model}
\end{figure*}

The spin-$\frac{1}{2}$ $J_{1}$--$J_{2}$--$J_{3}$ model on the honeycomb
lattice, or special cases of it (e.g., when $J_{3}=J_{2}$ or $J_{3}=0$) have been intensively studied 
by many authors (see, e.g.,
Refs.~[\onlinecite{Rastelli:1979,Mattson:1994,Fouet:2001,Mulder:2010,
Cabra:2011,Ganesh:2011,Clark:2011,Reuther:2011,DJJF:2011_honeycomb,Albuquerque:2011,
Mosadeq:2011,Oitmaa:2011,Mezzacapo:2012,PHYLi:2012_honeycomb_J1neg,
PHYLi:2012_honeyJ1-J2,PHYLi:2012_Honeycomb_J2neg}]
and references cited therein).  The Hamiltonian of the model is
%%%%%%%%%%%%%%%%%%%%
\begin{equation}
H = J_{1}\sum_{\langle i,j \rangle} \mathbf{s}_{i}\cdot\mathbf{s}_{j} + 
J_{2}\sum_{\langle\langle i,k \rangle\rangle} \mathbf{s}_{i}\cdot\mathbf{s}_{k} + 
J_{3}\sum_{\langle\langle\langle i,l \rangle\rangle\rangle} \mathbf{s}_{i}\cdot\mathbf{s}_{l}\,,
\label{eq1}
\end{equation}
%%%%%%%%%%%%%%%
where $i$ runs over all lattice sites, and where $j$, $k$, and $l$ run over all NN sites, 
all NNN sites, and all NNNN sites to $i$, respectively, counting each bond 
once and once only.  Each site $i$ of the lattice 
carries a spin-$s$ particle represented by an SU(2) spin operator 
${\bf s}_{i}=(s^{x}_{i},s^{y}_{i},s^{z}_{i})$.  We restrict ourselves here
to the case $s=\frac{1}{2}$.
The lattice and the exchange bonds are illustrated in Fig.~\ref{model}.
%%%%%%%%%%%%%%%%%%%%%%%%%%%%%%%%%%%%%%%%%%%%%
%%%%%%%%%%%%%%%%%%%%%%%%%%%%%%%%%%%%%%%%%%%%%%%%

Before discussing the $s=\frac{1}{2}$ version of the model that is the topic of
the present paper it is useful to consider first the classical limit (i.e., $s \rightarrow \infty$).
Thus the $J_{1}$--$J_{2}$--$J_{3}$ model on the honeycomb lattice has six
classical GS phases in the case where $J_{1}>0$ (as considered here) and where $J_2$
and $J_{3}$ are arbitrary (i.e., can take either sign).\cite{Rastelli:1979,Fouet:2001}  
The six phases comprise three collinear AFM phases, the FM phase, plus 
two different helical phases (and see, e.g., Fig. 2 of
Ref.~\onlinecite{Fouet:2001}).  The three AFM phases are the N\'{e}el (N) phase,
the striped (S) phase and the anti-N\'{e}el (aN) phase as shown in 
Figs.~\ref{model}(a), (b), and (d) respectively.  The S, aN, and N
states have, correspondingly, 1, 2, and all 3 NN spins to a given spin
antiparallel to it.  Similarly, if we consider the sites of the
honeycomb lattice as comprising a set of parallel sawtooth (or zigzag)
chains (in any one of the three equivalent directions), the S state
consists of alternating up-spin and down-spin FM chains, whereas the aN state consists of AFM
chains in which NN spins on adjacent chains are parallel.  

Although at $T=0$
there exists an infinite family of non-coplanar states degenerate in
energy with respect to each of the S and aN states, both
thermal and quantum fluctuations \cite{Fouet:2001} favor the collinear
configurations.  When $J_{3}>0$ the spiral state shown in
Fig.~\ref{model}(c) is the stable classical GS phase in some region of the parameter space.  
This state is characterized by a pitch vector perpendicular to one of the three equivalent $J_1$ bond directions, and a
single spiral angle defined so that as we move along the parallel sawtooth chains [drawn
in the horizontal direction in Fig.~\ref{model}(c)] the spin angle increases by $\pi + \phi$ from one site
to the next, and with NN spins on adjacent chains antiparallel.
The classical GS
energy for this spiral state is minimized when the pitch angle takes the value $\phi =
\cos ^{-1} [\frac{1}{4}(J_{1}-2J_{2})/(J_{2}-J_{3})]$.  The corresponding minimum value for the GS
energy per spin is then given as
%%%%%%%%%%%%%%%%%%%%
\begin{equation}
\frac{E^{\rm cl}_{\rm spiral}}{N} = \frac{s^{2}}{2}\left (-J_{1} - 2J_{2} + J_{3} - \frac{1}{4} 
\frac{(J_{1} - 2J_{2})^{2}}{(J_{2} - J_{3})} \right ). 
\end{equation}
%%%%%%%%%%%%%%%%

When $\phi \rightarrow 0$ this spiral state simply becomes the collinear N state with a
corresponding energy per spin given by
%%%%%%%%%%%%%%%%%%%%
\begin{equation}
\frac{E^{\rm cl}_{\rm N}}{N} = \frac{s^{2}}{2}\left (-3J_{1} + 6J_{2} - 3J_{3}\right). 
\end{equation}
%%%%%%%%%%%%%%%%%%%%
Clearly, the phase transition between this spiral state and the N state is of a continuous 
nature and the corresponding phase boundary is given by the equation
$y=\frac{3}{2}x-\frac{1}{4}$, for $\frac{1}{6}<x<\frac{1}{2}$, where $y\equiv J_{3}/J_{1}$ 
and $x\equiv J_{2}/J_{1}$.  

Similarly, when $\phi \rightarrow \pi$ this spiral state becomes
the collinear S state with a corresponding GS energy per spin given by
\begin{equation}
\frac{E^{\rm cl}_{\rm S}}{N} = \frac{s^{2}}{2}\left (J_{1} - 2J_{2} - 3J_{3}\right). 
\end{equation}
%%%%%%%%%%%%%%%%%%%%
The classical spiral and S states undergo a continuous phase transition along their common
phase boundary $y=\frac{1}{2}x+\frac{1}{4}$, for
$x>\frac{1}{2}$.  Furthermore there is a first-order phase
transition between the collinear N and S states along the boundary line $x=\frac{1}{2}$, for
$y>\frac{1}{2}$.  These three phases (N, S, and spiral) meet at the
tricritical point $(x,y) = (\frac{1}{2},\frac{1}{2})$.  We note too that as $x
\rightarrow \infty$ (for a fixed finite value of $y$) the spiral
pitch angle $\phi \rightarrow \frac{2}{3} \pi$.  Thus in this limiting case the classical model
simply becomes two HAFMs on weakly connected
interpenetrating triangular lattices, with the classical triangular-lattice ordering of
NN spins oriented at an angle $\frac{2}{3} \pi$ to each other on each
sublattice.

When $y>0$ (and $J_{1}>0$) the above three states are the only classical GS phases.  When $y<0$ the N state
persists in a region bounded by the same boundary line as above, 
$y=\frac{3}{2}x-\frac{1}{4}$, for $-\frac{1}{2}<x<\frac{1}{6}$, on which it continuously
meets a second spiral state, and by the boundary line $y=-1$, for $x<-\frac{1}{2}$, at which 
it undergoes a first-order transition to the FM state, which itself is the stable GS phase
in the region $x<-\frac{1}{2}$ and $y<-1$.   Another collinear AFM state, the 
aN state shown in Fig.~\ref{model}(d), with a GS energy per spin given by
%%%%%%%%%%%%%%%%%%%%%%
\begin{equation}
\frac{E^{\rm cl}_{\rm aN}}{N} = \frac{s^{2}}{2}\left (-J_{1} - 2J_{2} + 3J_{3}\right), 
\end{equation}
%%%%%%%%%%%%%%%%%%%%
becomes the stable GS phase in the region $x>\frac{1}{2}$, for 
$y < \frac {1}{2}\{x -[x^{2}+2(x-\frac{1}{2})^{2}]^{1/2} \}$.  
On the boundary it
undergoes a first-order transition to the spiral state shown in Fig.~\ref{model}(c).

Finally, for $\frac{1}{6}<x<\frac{1}{2}$ the spiral state shown in Fig.~\ref{model}(c) meets
a second spiral GS phase on the boundary line $y=0$, along which there is a 
first-order transition between the two spiral states.  This second spiral
phase is characterized by a pitch vector that is parallel to one of the $J_1$ bond directions and by
two spiral angles $\alpha$ and $\beta$.  Within a unit cell the spin directions deviate from those in
the N state by an angle $\alpha$, and as one advances from one unit cell to the next there is
a twist by an angle $\beta$.  Both of the two pitch angles of this second spiral phase smoothly approach
the value zero along the above boundary with the N state, and the value $\pi$ along a
second boundary curve that joins the points $(x,y) = (-\frac{1}{2},-1)$ 
and $(\frac{1}{2},0)$, on which it meets the aN state.  Both transitions are continuous in nature.
This second spiral phase meets the three collinear states N, aN, and FM at the tetracritical
point $(x,y) = (-\frac{1}{2},-1)$.

In this paper we further our study of the spin-$\frac{1}{2}$ model of Eq.~(\ref{eq1}) on the honeycomb
lattice, restricting ourselves to the case where all of the bonds are 
antiferromagnetic in nature.  Thus, henceforth we set $J_{1} \equiv 1$ to set the overall 
energy scale, and we work here within the parameter space window $J_{2},J_{3} \in [0,1]$.

Recently we have used the CCM to study the special $J_{1}$--$J_{2}$ case of this model 
(and in this parameter window with $J_{1}>0$ and $J_{2} \equiv xJ_{1}>0$) in which 
$J_{3}=0$.\cite{PHYLi:2012_honeyJ1-J2}  We found a paramagnetic plaquette valence-bond
crystalline (PVBC) phase for
$x_{c_1}<x<x_{c_2}$, where $x_{c_1} \approx 0.207
\pm 0.003$ and $x_{c_2} \approx 0.385 \pm 0.010$.  We found that the transition
at $x_{c_1}$ to the N phase appeared to be of a continuous
deconfined type (although we could not exclude a very narrow
intermediate phase in the range $0.21 \lesssim x \lesssim 0.24$),
while that at $x_{c_2}$ to the aN phase appeared to be of first-order type.  As we noted above 
the aN phase exists in the classical
version of the $J_{1}$--$J_{2}$ model only at the isolated and highly degenerate
critical point $x = \frac{1}{2}$.  The spiral phases that are
present classically for all values $x > \frac{1}{6}$ were found to be absent
for all $x \lesssim 1$.  

We have also separately used the CCM to study the spin-$\frac{1}{2}$ model of Eq.~(\ref{eq1}) on the honeycomb
lattice in the special case where $J_{3}=J_{2}\equiv \kappa J_{1}>0$ and $J_{1}>0$ (i.e. along
the line $y=x \equiv \kappa$).\cite{DJJF:2011_honeycomb}  We also found a PVBC phase in this case for
$\kappa_{c_1}<\kappa<\kappa_{c_2}$, where $\kappa_{c_1} \approx 0.47$ and
$\kappa_{c_2} \approx 0.60$.  Once again, the evidence favored the transition at $\kappa_{c_1}$ to the N
phase to be of a continuous deconfined type, while that at $\kappa_{c_2}$ to the S phase appeared
to be first-order in nature.

In order to shed more light on the
$J_{1}$--$J_{2}$--$J_{3}$ HAFM model (i.e., with $J_{1}>0$) on the honeycomb lattice, we now extend
its study to map out its entire phase diagram in the parameter regime $x,y \in [0,1]$.  In view of
the proven success of using the CCM on the above special cases of this model, we continue
to utilize it, and in Sec.~\ref{CCM} we accordingly first briefly outline the method as it
is applied here.

\section{THE CCM FORMALISM}
\label{CCM}
The CCM is a widely used microscopic many-body technique.  It is
has been demonstrated to be particularly efficient and very accurate in handling 
a wide variety of highly frustrated quantum magnets.  Such 
frustrated systems are notoriously challenging at the theoretical level.
Only a very limited number of established numerical methods exist for their
accurate treatment, and other recent and very promising approaches such as those based
on projected entangled pair states\cite{peps} have not yet been sufficiently
widely tested to be able to evaluate properly their accuracy and efficacy.

Of the other well established and widely used techniques we note that
exact diagonalization (ED) techniques, which involve the finite-size extrapolation 
of numerical exact data for finite-lattice systems, are much more
challenging for the present honeycomb-lattice model than for comparable
square-lattice models, to which they have been very efficiently and accurately
applied (see, e.g., Refs.~[\onlinecite{schulz,Richter:2010_ED,Reuther:2011_J1J2J3mod}]).
The reasons include the facts that for the honeycomb lattice the unit cell
now contains two sites, and that there exist relatively fewer finite-sized 
lattices (than in the square-lattice case) that
are small enough for ED techniques to be used but which also contain the
full point-group symmetry.\cite{Fouet:2001}  Also, quantum Monte
Carlo (QMC) methods are severely restricted in the presence of
frustration by the well-known ``minus-sign problem''.  

By contrast the CCM, when evaluated to high orders in one of its systematic 
approximation hierarchies, as described below, has been proven through a huge
variety of applications to provide a powerful tool both to determine with good accuracy
the positions of quantum critical points, 
\cite{DJJF:2011_honeycomb,PHYLi:2012_honeycomb_J1neg,
PHYLi:2012_honeyJ1-J2,PHYLi:2012_Honeycomb_J2neg,
Reuther:2011_J1J2J3mod,Kr:2000,rachid05,schmalfuss,Bi:2008_PRB,Bi:2008_JPCM,
darradi08,Bishop:2009,richter10,UJack_ccm,Bishop:2012_checkerboard}
and to classify the nature of any QP phases in the system.\cite{DJJF:2011_honeycomb,
PHYLi:2012_honeyJ1-J2,darradi08,Bishop:2012_checkerboard}  As noted above, we have
also used the CCM very successfully in some previous applications to honeycomb-lattice 
models,\cite{DJJF:2011_honeycomb,PHYLi:2012_honeycomb_J1neg,
PHYLi:2012_honeyJ1-J2,PHYLi:2012_Honeycomb_J2neg} and for all these reasons
we now employ it again here.

The CCM is a size-extensive method, in which the limit $N \rightarrow \infty$,
where $N$ is the number of lattice spins,
may automatically be imposed from the outset.  The many-body system under study is assumed to have 
exact ket and bra GS energy eigenvectors, $|\Psi\rangle$ and $\langle \tilde{\Psi}|$ 
respectively, which satisfy the corresponding Schr\"{o}dinger equations,
\begin{equation}
H|\Psi\rangle=E|\Psi\rangle\,, \quad \langle \tilde{\Psi}|H=E\langle\tilde{\Psi}| \,,
\end{equation}
and which are chosen to have the normalization $\langle \tilde{\Psi}|\Psi \rangle = 1$, i.e.,
$\langle \tilde{\Psi}| = \langle \Psi|/\sqrt{\langle \Psi|\Psi\rangle}$.
The quantum correlations present in the exact ground state are expressed
systematically in the CCM with respect to some suitable normalized model (or reference)
state, $|\Phi\rangle$.\cite{ccm2,j1j2_square_ccm1,ccm3}  It is common practice to choose
simple quasiclassical states as CCM reference states, although other
choices are certainly possible.  In this study we choose various
classical model states as our CCM model states, namely: (a) the N\'{e}el,
(b) the striped, (c) the spiral, and (d) the anti-N\'{e}el states
shown in Fig.~\ref{model}, as we discuss below.  

The model state $|\Phi\rangle$ is required to be a fiducial vector in the sense that
all possible ket states in the many-body Hilbert space can be obtained by acting
on it with an appropriate linear combination of mutually commuting many-body creation operators,
$C^{+}_{I}$, which may be defined with respect to the model state.
The operators $C^{+}_{I} \equiv (C^{-}_{I})^{\dagger}$, with 
$C_{0}^{\dagger} \equiv 1$, thus have the property that 
$\langle \Phi|C^{+}_{I} = 0 = C_{I}^{-}|\Phi\rangle = 0$\,; $\forall I \neq 0$.  
The CCM parametrizations of the exact ket and bra GS wave functions are
given in terms of the usual exponentiated forms,
\begin{equation}
|\Psi\rangle = {\rm e}^{S}|\Phi\rangle\,, \quad \langle\tilde{\Psi}| = \langle \Phi|\tilde{S}{\rm e}^{-S} \,,
\end{equation}
where the CCM correlation operators $S$ and $\tilde{S}$ are themselves expressed
as generalized multiconfigurational creation and destruction operators respectively,
\begin{equation}
S=\sum_{i}{\cal S}_{I} C^{+}_{I}\,, \quad \tilde{S}_{I}=1 +\sum_{i} {\tilde{\cal S}}_{i} C^{-}_{I} \,\,,  
\forall I \neq 0 \,.
\end{equation} 
Clearly these parametrizations satisfy the normalization relations
$\langle \tilde{\Psi}|\Psi \rangle = \langle\Phi|\Psi\rangle = \langle\Phi|\Phi\rangle \equiv 1$.

The set of correlation coefficients (${\cal S}_{I}$, $\tilde{\cal S}_{I}$) is
now determined by requiring the energy expectation value
$\bar{H}\equiv\langle\tilde{\Psi}|H|\Psi\rangle$ to be a minimum with respect to
each of the correlation coefficients themselves.  This will result in the coupled sets of 
equations $\langle \Phi |C^{-}_{I}{\rm
  e}^{-S}H{\rm e}^{S}|\Phi\rangle=0$ and
$\langle\Phi|\tilde{S}({\rm e}^{-S}H{\rm e}^{S} -
E)C^{+}_{I}|\Phi\rangle=0$; $\forall I \neq 0$, which we normally solve for 
the correlation coefficients (${\cal S}_{I}$, $\tilde{\cal S}_{I}$) using
parallel computing routines once the specific truncation scheme is specified,
as described further below.

In order to treat each lattice site in the spin system on an equal basis it is
extremely convenient to rotate the local spin-axes on each site in such a
way that all the  spins of each CCM reference state used point along the negative $z$-direction.  
Such rotations in spin space are obviously canonical transformations that have 
no effect on the fundamental SU(2) commutation relations.  The spins of our
system are then represented entirely by these locally defined spin
coordinate frames.  The multispin creation operators may be written as
linear sums of products of the individual spin raising operators
$s^{+}_{k} \equiv s^{x}_{k} + is^{y}_{k}$, i.e., $C^{+}_{I} \equiv
s^{+}_{k_{1}}s^{+}_{k_{2}} \cdots s^{+}_{k_{n}}$.  After calculation of
the correlation coefficients (${\cal S}_{I}$, $\tilde{\cal S}_{I}$),
we can then calculate the GS energy using $E=\langle\Phi|{\rm
  e}^{-S}H{\rm e}^{S}|\Phi\rangle$, and the magnetic order parameter,
which is defined to be the average local on-site magnetization,
$M \equiv -\frac{1}{N}\langle\tilde{\Psi}| \sum^{N}_{i=1} s^{z}_{i}|
\Psi\rangle$, with respect to the local rotated spin coordinates described above.

If we include all possible multispin configurations for the calculation of the 
correlation coefficients (${\cal S}_{I}$, $\tilde{\cal S}_{I}$), then the CCM formalism becomes exact.  
Of course in practice one needs to truncate the set, and there are several well-developed and 
systematically improvable truncation hierarchies that have been extremely
widely tested by now.  For spin-$\frac{1}{2}$ systems we usually
use the well-established localized LSUB$m$ truncation scheme where we
keep at a given truncation level specified by the truncation index $m$ only all of
those multi-spin configurations which may be defined over all possible lattice
animals (or polyominos) of size $m$ on the lattice.  A lattice animal (or polyomino) of
size $m$ is defined as a set of $m$ contiguous sites in the usual graph-theoretic sense 
where every site is adjacent (in the nearest-neighbor sense) to
at least one other site.  The method of solving
for higher orders of LSUB$m$ approximations is well documented in
Refs.~[\onlinecite{ccm2,ccm3}], to which the interested reader is referred
for further details.  

Table \ref{table_FundConfig} shows the number $N_f$ of fundamental
configurations that are inequivalent after all space and point-group symmetries
of both the Hamiltonian and the model state have been taken into account, 
for each of the N\'{e}el, striped, spiral, and anti-N\'{e}el model
states of the spin-$\frac{1}{2}$ $J_{1}$--$J_{2}$--$J_{3}$ model on
the honeycomb lattice.
%%%%%%%%%%%%%%%%%%%%%%%%%%%%%%%%%%%
\begin{table}
  \caption{Number of fundamental configurations, $N_{f}$, for the 
spin-$\frac{1}{2}$ $J_{1}$--$J_{2}$--$J_{3}$ model ($J_{1}=1$) on the honeycomb lattice, 
using the N\'{e}el, striped, anti-N\'{e}el, and spiral states.}
\vskip0.2cm
\begin{tabular}{|c|c|c|c|c|} \hline\hline
Method & \multicolumn{4}{|c|}{$N_{f}$} \\ \cline{2-5}    
&      N\'{e}el & striped  & anti-N\'{e}el   & spiral    \\ \hline
LSUB4 & 5       & 9        & 9     & 66 \\
LSUB6 & 40      & 113       & 85    &1080  \\ 
LSUB8 & 427     & 1750      &1101   &18986  \\
LSUB10 & 6237   & 28805     &17207  & 347287 \\   \hline\hline     
\end{tabular} 
\label{table_FundConfig}
\end{table}
%%%%%%%%%%%%%%%%%%%%%%%%%%%%%%%%%%
We note that the number $N_f$ of such independent spin configurations taken
into account in the CCM correlation operators $S$ and $\tilde{S}$ increases rapidly with the
truncation index $m$.  Clearly the number of independent configurations
is smaller for states such as the N\'{e}el state that have a higher degree
of point-group symmetry, and for which we can utilize conservation laws
such as $s^{z}_{T}=0$, where ${\bf s}_{T} \equiv \sum_{i=1}^{N} {\bf s}_{i}$ 
is the total spin operator referred to the global spin coordinates.

Clearly the spiral state has the largest number $N_f$, for a given level
of LSUB$m$ approximation, from among our four model states, and for this state we
are limited to values $m \leq 10$ even with the use of massively
parallel computing to derive and solve the corresponding coupled sets of 
CCM bra- and ket-state equations.\cite{ccm}  For the spiral state we note too
that we have the additional computational cost that the pitch angle $\phi$ at
a given LSUB$m$ level must be chosen to minimize the corresponding
estimate for the GS energy.  For specified values of each of the exchange
parameters $J_{2}$ and $J_{3}$ (with $J_{1} \equiv 1$) a typical computational
run for the spiral phase at the LSUB$10$ level typically requires about 
5$\,$h computing time using 3000 processors simultaneously.

Although the CCM works from the outset in the limit $N \to \infty$ of
an infinite number of spins, and hence the need for any finite-size 
scaling is obviated, we do still need to extrapolate the
LSUB$m$ data to reach results in the exact $m \rightarrow \infty$ limit.  Although there are no known
exact extrapolation rules, by now there exists a wealth of empirical experience in extrapolating the GS
energy, $E$, and the magnetic order parameter (i.e., the average local on-site
magnetization), $M$. 
For the GS energy per spin, $E/N$, a well-established and very accurate
extrapolation ansatz (see, e.g., Refs.
[\onlinecite{Reuther:2011_J1J2J3mod,Kr:2000,rachid05,schmalfuss,
Bi:2008_PRB,Bi:2008_JPCM,darradi08,Bishop:2009,richter10,UJack_ccm,
Bishop:2012_checkerboard,ccm3}]) is
%%%%%%%%%%%%
\begin{equation}
E(m)/N = a_{0}+a_{1}m^{-2}+a_{2}m^{-4}\,,     \label{E_extrapo}
\end{equation}
whereas for the magnetic order parameter, $M$, we use different schemes
depending on different circumstances, specifically on whether the system
is highly frustrated or not.  Thus, for systems with a GS
order-disorder transition or with a considerable degree of frustration, 
such as is the case for the present model, we use (see, e.g.,
Refs. [\onlinecite{DJJF:2011_honeycomb,PHYLi:2012_honeycomb_J1neg,
PHYLi:2012_honeyJ1-J2,PHYLi:2012_Honeycomb_J2neg,
Reuther:2011_J1J2J3mod,Bi:2008_PRB,Bi:2008_JPCM,darradi08,richter10,
Bishop:2012_checkerboard}])
%%%%%%%%%%%%%
\begin{equation}
M(m) = c_{0}+c_{1}m^{-1/2}+c_{2}m^{-3/2}\,.    \label{M_extrapo_frustrated}
\end{equation}
%%%%%%%%%%%%%%%

When we have only three data points to fit to an extrapolation formula, 
such as will sometimes occur here, specifically for the spiral phase, a two-term
extrapolation fit can easily be
preferable in practice to a three-term fit.  This is particularly the case when one of the data
points is either far from the limiting case or when it does not
represent all of the features of the system as well as the remaining,
more accurate points.  In such cases we sometimes use the alternative
simpler forms,
%%%%%%%%%%%%%
\begin{equation}
E(m)/N = b_{0}+b_{1}m^{-2}\,,     \label{E_extrapo_linearFit}
\end{equation}
%%%%%%%%%%%%%%%%%%%%
and
%%%%%%%%%%%%%%%
\begin{equation}
M(m) = d_{0}+d_{1}m^{-1/2}\,,   \label{M_extrapo_frustrated_linearFit}
\end{equation}
%%%%%%%%%%%%%%%
instead of their counterparts in Eqs.~(\ref{E_extrapo}) 
and (\ref{M_extrapo_frustrated}), respectively.  

Finally we note that since the hexagon is an important structural element of the honeycomb
lattice, it is preferable for the extrapolations to use only
LSUB$m$ data with $m \geq 6$, wherever possible.  However, especially for the spiral
phase that is particularly costly of computational resource, as we explain below,
we sometimes need to include LSUB4 results in the extrapolations.
Under such circumstances, however, we always perform a sensitivity analysis, by doing
some LSUB$m$ runs with higher values of $m$ for a few indicative points only,
as we discuss in more detail in Sec.~\ref{results}.  

\section{PREVIEW OF THE PHASE DIAGRAM}
\label{preview}
%%%%%%%%%%%%%%%%%%
 \begin{figure}[!tb]
\includegraphics[width=6cm,angle=270]{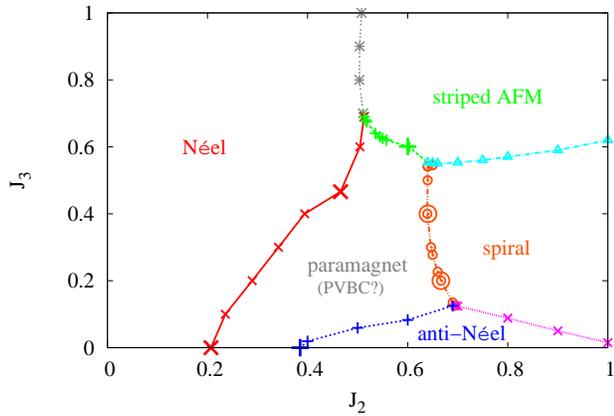}
\caption{ (Color online) Phase diagram of the spin-$\frac{1}{2}$
  $J_{1}$--$J_{2}$--$J_{3}$ model on the honeycomb lattice (with
  $J_{1} \equiv 1$), in the parameter window $J_{2},J_{3} \in [0,1]$.
  The five regions correspond to four quasiclassical phases with (a)
  AFM N\'{e}el (N) order as shown in Fig.~\ref{model}(a), (b)
  collinear AFM striped (S) order as shown in Fig.~\ref{model}(b), (c)
  spiral order as shown in Fig.~\ref{model}(c), (d) AFM anti-N\'{e}el
  (aN) order as shown in Fig.~\ref{model}(d), plus (e) a magnetically
  disordered, or quantum paramagnetic (QP), phase that exhibits
  plaquette valence-bond crystalline (PVBC) order on at least part of
  the boundary region (and see below).  The first-order phase
  transition boundary between the N and S phases, marked by the (grey)
  convolution (eight-pointed star, \plusontimes) symbols is found from
  points at which the curves for the magnetic order parameter $M$ of
  the two phases cross; the first-order phase transition boundary
  between the S and QP phases, marked by (green) plus ($+$) symbols,
  is found from points at which $M \to 0$ for the S phase; the
  first-order phase transition boundary between the S and spiral
  phases, marked by (cyan) open triangle ($\triangle$) symbols, is
  found from points at which the curves for the magnetic order
  parameter $M$ of the two phases cross; the phase transition boundary between
  the spiral and QP phases, marked by (orange) open circle
  ($\bigcirc$) symbols (of two sizes, see main text in Sec.~\ref{results_neel_spiral}), is found from points at which $M \to 0$ for
  the spiral phase; the first-order phase transition boundary
  between the spiral and aN states, marked by (magenta) times
  ($\times$) symbols, is
  found from points at which the curves for the magnetic order
  parameter $M$ of the two phases cross; the phase transition boundary between
  the aN and QP phases, marked by (blue) plus
  ($+$) symbols, is found from points at which $M \to 0$ for
  the aN phase; and the phase
  transition boundary between the N and QP states, marked by (red)
  times ($\times$) symbols, which is probably of continuous
  (second-order, and possibly of a deconfined) nature, is found from
  points at which $M \to 0$ for the N phase.  Points marked by the
  larger (red) times ($\times$) and (green and blue) plus ($+$) symbols are found to be
  infinitely susceptible to PVBC order, and hence the QP state at
  these points is PVBC in nature.} 
\label{phase}
\end{figure}
%%%%%%%%%%
Before discussing our results in detail it is perhaps useful to summarize 
our main findings first, and for that purpose we show in Fig.~\ref{phase} the phase
diagram for the frustrated spin-$\frac{1}{2}$ model of Eq.~(\ref{eq1}) on
the honeycomb lattice, in the case where all the bonds are antiferromagnetic
in nature (i.e., $J_{n} > 0,\, n=1,2,3$).  Furthermore, we set $J_{1} \equiv 1$ and restrict
ourselves to the window $0 \leq J_m \leq 1, m=2,3$.  Henceforth we denote
$x \equiv J_{2}/J_{1}$, $y \equiv J_{3}/J_{2}$.  The actual phase boundaries
are determined from a variety of information that emerges from our CCM
calculations, as we now describe briefly and with further details given in
Sec.~\ref{results}.  

As we have already noted, we have previously studied this model for the two special
cases with $J_{3}=J_{2}$ in Ref.~[\onlinecite{DJJF:2011_honeycomb}], and with 
$J_{3}=0$ in Ref.~[\onlinecite{PHYLi:2012_honeyJ1-J2}], and the corresponding CCM 
results from those papers are included in Fig.~\ref{phase}.  Firstly, along the
line $J_{3}=J_{2}$ (i.e., when $y = x \equiv \kappa$) we found\cite{DJJF:2011_honeycomb} 
that the system has quasiclassical AFM
N\'{e}el (N) order for $\kappa < \kappa_{c_1} \approx 0.47$, quasiclassical AFM striped (S) order
for $\kappa > \kappa_{c_2} \approx 0.60$, and a quantum paramagnetic phase separating the N and S phases for
$\kappa_{c_1} < \kappa < \kappa_{c_2}$.  By studying the susceptibility of the N and S states to
hexagonal plaquette valence-bond crystal (PVBC) ordering, we found that the most likely scenario
was that the intervening state had PVBC order over the entire range
$\kappa_{c_1} < \kappa < \kappa_{c_2}$.  The transition at $\kappa = \kappa_{c_2}$ between the
PVBC and S GS phases was seen to be of first-order type, while that at
$\kappa = \kappa_{c_1}$ between the N and PVBC GS phases appeared to be a continuous one.
Since the N and PVBC phases break different symmetries our results favored the transition point
between them at $\kappa = \kappa_{c_1}$ to be a deconfined quantum critical point (QCP).
The QCPs at $y = x = \kappa_{c_1}$ and at $y = x = \kappa_{c_2}$ are clearly shown in
Fig.~\ref{phase} with the larger (red) times ($\times$) and the larger (green) plus ($+$)
symbols respectively. 

Secondly, in a separate study along the line $y=0$, we
found\cite{PHYLi:2012_honeyJ1-J2} that the system has the
quasiclassical N state as its GS phase for $x < x_{c_1} \approx 0.21$,
the quasiclassical anti-N\'{e}el (aN) state as its GS phase for $x >
x_{c_2} \approx 0.39$, and again a quantum paramagnetic (QP) phase separating the N
and aN phases for $x_{c_1} < x < x_{c_2}$.  Similar CCM calculations
of the susceptibility of the N and aN phases to PVBC order led again
to the conclusion that the transition between the PVBC and aN phases
was of first-order type, while the likely scenario for the transition
between the N and PVBC phases is again that it is of the continuous
deconfined type.  Nevertheless, due to the difficulty in determining
the lower critical value of $x$ at which PVBC order is established as
accurately as we determined the value $x = x_{c_1}$ at which N\'{e}el
order is destabilized, we could not exclude a second scenario in which
the transition between the N and PVBC phases proceeds via an
intervening phase (possibly even of an exotic spin-liquid variety) in
the very narrow window $0.21 \lesssim x \lesssim 0.24$.  Again, the
QCPs at $(x,y) = (x_{c_1},0)$ and $(x_{c_2},0)$ are clearly shown in
Fig.~\ref{phase} with the larger (red) times ($\times$) and the
(blue) plus ($+$) symbols respectively.

We also showed previously,\cite{PHYLi:2012_honeyJ1-J2} by a comparison of the GS energies
of the spiral and aN phases calculated separately with the CCM,
that the spiral phases that are present classically (i.e., for
the case where the spin quantum number $s \to \infty$) in the case $y=0$ for all values $x > \frac{1}{6}$
are absent for all values $x \lesssim 1$.  The actual phase boundary between the spiral and aN
phases shown in Fig.~\ref{phase} is now calculated in the present 
paper, as described below.

Based on our previous findings for the GS phases of these two special cases when
(a) $J_{3} = J_{2}$ and (b) $J_{3} = 0$. we have now performed a series of CCM calculations
based on the N, S, aN, and spiral states as model states, for a variety of cuts in
the phase diagram at both constant values of $J_3$ and constant values of $J_2$.  For example,
the phase boundary between the N and the S phases is obtained, as explained more fully in
Sec.~\ref{results_neel_striped}, from our extrapolated ($m \to \infty$) LSUB$m$ results for the order 
parameter $M$ (namely, the average onsite magnetization) of the two phases, for a variety
of constant $J_3$ cuts.  We find that for values of $y \equiv J_{3}/J_{1} \gtrsim 0.69$ the two
magnetization curves meet at a (positive) nonzero value, indicative of a direct first-order
transition between the states.  These points are shown in Fig.~\ref{phase} by the (grey)
convolution (eight-pointed star, \plusontimes) symbols.

For the value $y \approx 0.69$ the two curves become zero at
precisely the same point, $x \approx 0.51$. 
Conversely, when $y \lesssim 0.69$, the order parameters 
of the N and the S phases both become zero at respective critical 
values of $x$ before the curves cross (when solutions 
exist for both phases), indicating the emergence of
a new phase separating them.  The corresponding points where the magnetic order parameters for N\'{e}el order and striped order
vanish are shown in Fig.~\ref{phase} by (red) times ($\times$) and (green) plus ($+$)
symbols respectively.
By continuity with our earlier results\cite{DJJF:2011_honeycomb} along the 
line $y=x$, we tentatively identify the intervening phase as the PVBC state.  The 
tricritical QCP between the N, S, and PVBC phases is thus identified as being
at $(x,y) \approx (0.51,0.69)$. We also note that for values of $y \lesssim 0.55$ no
solution for the S phase exists with $M>0$ for any value of $x$, giving preliminary indications of a new phase boundary
between the S state and another phase that we identify as a spiral phase.

By comparing the order parameters for the S and spiral phases at various constant $J_2$ cuts
we find that for values of $x \gtrsim 0.66$ the two curves meet
at a (positive) nonzero value, once again indicative of a direct 
transition between the states.  These points are shown in Fig.~\ref{phase} as (cyan) open triangle ($\triangle$)
symbols.  For the value $x \approx 0.66$ the two curves become zero at
the same point $y \approx 0.55$.  Then, for values $x \lesssim 0.66$ the order parameters of the S and spiral 
phases both become zero at respective critical values of $y$ before the curves cross.  Once again this
indicates a phase separating the S and spiral phases for values of $x \lesssim 0.66$ (down to
a lower value of $x \approx 0.635$ below which the spiral phase ceases to exist for any value of $y$), which we similarly 
identify tentatively as the PVBC phase.  

In that very narrow window $0.635 \lesssim x \lesssim 0.66$, which is
almost certainly an artifice of our approximations, we denote in Fig.~\ref{phase} the points where the magnetic order parameter
vanishes ($M \to 0$) for the striped and spiral states by (cyan) open triangle ($\triangle$)
and (orange) open circle ($\bigcirc$) symbols respectively.
We argue in Sec.~\ref{results_spiral_striped} that these results are
consistent with the existence of a second tricritical QCP at
$(x,y) \approx (0.65,0.55)$ between the S, spiral, and (tentatively) PVBC phases.  
The remainder of the phase boundary between the spiral and PVBC
states is similarly identified by the vanishing of the magnetic order parameter of the spiral phase, and these points
are again shown in Fig.~\ref{phase} as (orange) open circle ($\bigcirc$) symbols.

Finally by comparing the energies of the aN and spiral phases we find
that for all values of the parameter $J_{2} \leq 1$ where the spiral
phase exists, the aN phase actually has a lower energy for values of
the parameter $J_3$ below a certain critical value, which itself
depends on $J_2$.  Similarly, by comparing the order parameters of these two phases at various
constant $J_2$ cuts, we find that for values of $x \gtrsim 0.69$ 
the two curves meet at a (positive) nonzero value, indicative once more of a direct
phase transition between the aN and spiral phases.  These points are shown in
the phase diagram of Fig.~\ref{phase} by (magenta) times ($\times$) symbols.

For the value $x \approx 0.69$ the two curves become zero at the same point $y \approx 0.12$. 
Conversely, for values $x \lesssim 0.69$ the order parameters of the aN and spiral 
phases both become zero at respective critical values of $y$ before the curves cross.  This is again
indicative of a phase separating the aN and spiral phases for values of $x \lesssim 0.69$ (down
to the lower value of $x \approx 0.635$ below which the spiral phase ceases to exist for any value of $y$), as
noted above.  This intermediate phase is again tentatively identified as having PVBC order.  In this way we 
identify a third tricritical QCP at $(x,y) \approx (0.69,0.12)$ between the spiral, aN and (tentatively) PVBC phases.  
Remaining points on the phase boundary between the aN and PVBC phases are then identified as the
points where the magnetic order parameter of the aN phase vanishes, and these are shown by
(blue) plus ($+$) symbols on the the phase diagram of Fig.~\ref{phase}.

In Sec.~\ref{results} we now describe in more detail how the various points in the phase diagram
of Fig.~\ref{phase} are obtained.  We also discuss the properties of the various phases that we have examined.

\section{RESULTS}
\label{results}
In this section, we present and discuss our CCM results for 
the spin-$\frac{1}{2}$ $J_{1}$--$J_{2}$--$J_{3}$ HAFM
on the honeycomb lattice, with all of the bond strengths positive
(i.e., antiferromagnetic in nature).  To set the overall energy scale
we put $J_{1} \equiv 1$, and we investigate the parameter space 
window $J_{2},J_{3} \in [0,1]$.  We use each of the N\'{e}el (N), 
collinear striped (S), spiral, and anti-N\'{e}el (aN) states shown
respectively in Figs.~\ref{model}(a)-(d) as CCM model states.

\subsection{N\'{e}el versus striped phases}
\label{results_neel_striped}
\begin{figure*}
\mbox{
\subfigure[]{\scalebox{0.3}{\includegraphics[angle=270]{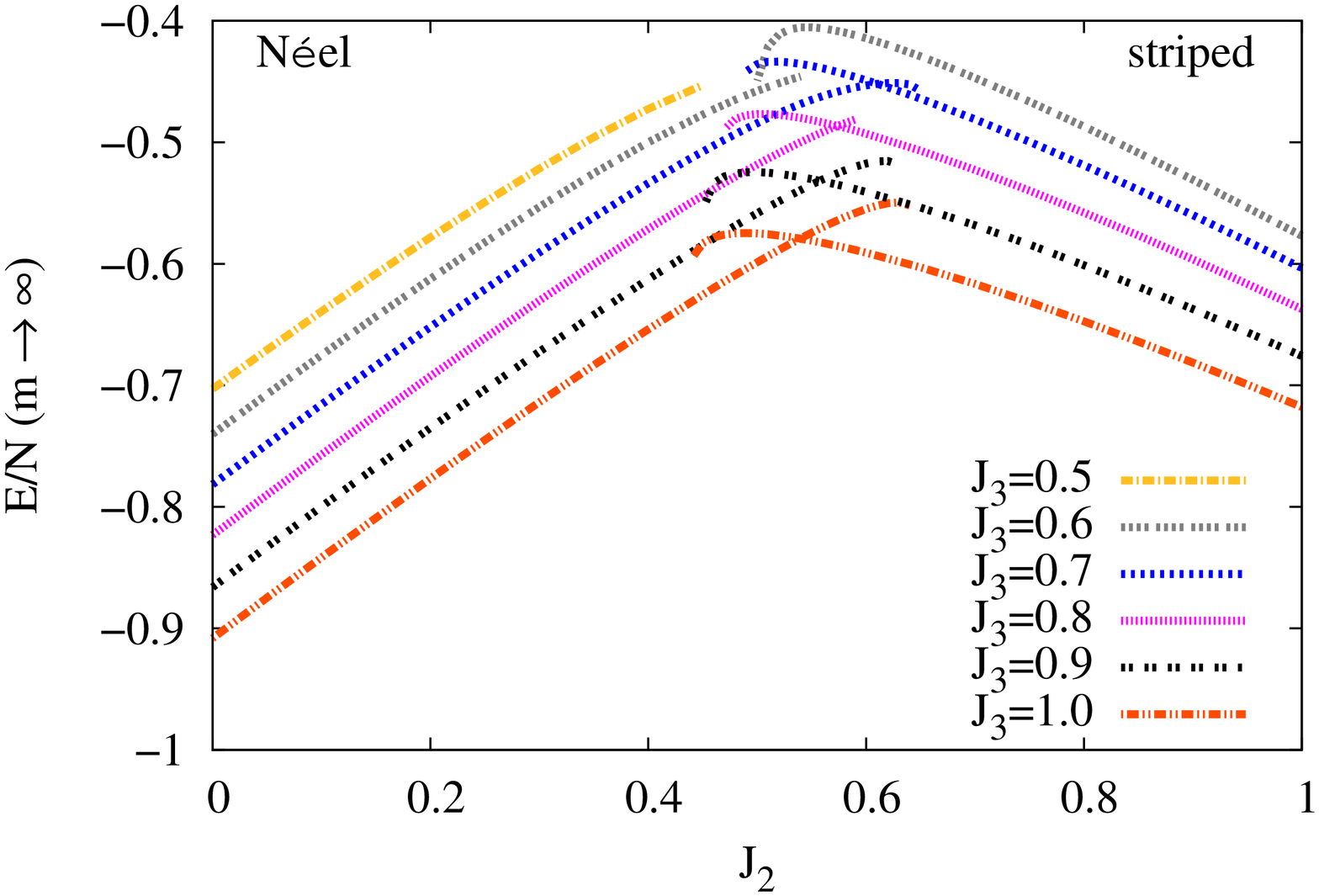}}}
\subfigure[]{\scalebox{0.3}{\includegraphics[angle=270]{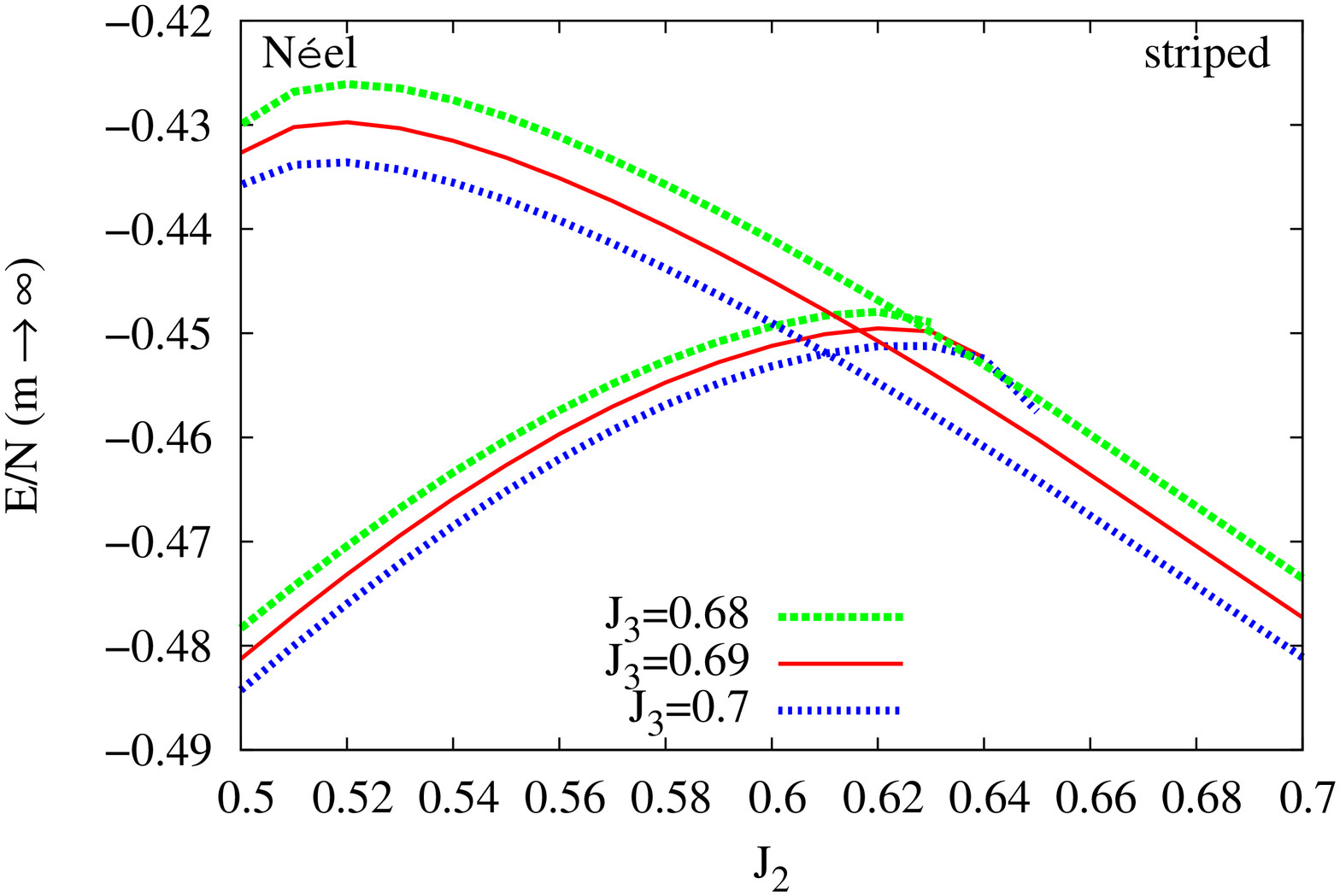}}}
}
\caption{ 
  (Color online) Extrapolated CCM LSUB$\infty$ results for the
  GS energy per spin, $E/N$, as a function of $J_{2}$, for various
  fixed values of $J_{3}$ in the range $0.5 \leq J_{3} \leq 1.0$, for
  the N\'{e}el and the striped states of the spin-$\frac{1}{2}$
  $J_{1}$--$J_{2}$--$J_{3}$ model on the honeycomb lattice (with
  $J_{1} \equiv 1$).  The extrapolated LSUB$m$ ($m \rightarrow
  \infty$) results are based on the extrapolation scheme of
  Eq.~(\ref{E_extrapo}) and the calculated results with
  $m=\{6,8,10\}$.
}
\label{E}
\end{figure*}

Figures \ref{E}(a) and \ref{E}(b)
%%%%%%%%%%%%%%%%%%%%%%%%%%%%%%%%%%%%
%%%%%%%%%%%%%%%%%%%%%%%%%%%%%%%%%%%%% 
show the extrapolated ($m \to \infty)$ CCM LSUB$m$ 
values for the GS energy per spin for the N\'{e}el (N) and striped (S) states as functions of $J_{2}$
for various fixed values of $J_{3}$ in the range $0.5 \leq J_{3} \leq 1.0$.  The
extrapolations have been performed using Eq.~(\ref{E_extrapo}) and the calculated LSUB$m$
results with $m=\{6,8,10\}$.  We observe that the energy curves cross for all
values of the parameter $J_{3} \gtrsim 0.68$, but for values $J_{3} \lesssim 0.68$ 
the curves do not cross.  This gives us a first indication of the emergence of 
an intermediate phase between the N and S states, over a finite range of
values of the $J_2$ parameter, below some critical value of the $J_3$ parameter.

We note that the extrapolations become more difficult in the
vicinity of this critical point, and consequently the actual values of $J_2$
at which the curves cross for fixed values of $J_3$ near the critical value
are more uncertain than those at larger values.  Furthermore, at the actual energy crossing
points very near the critical point the corresponding values of the magnetic order 
parameter (i.e., the average onsite magnetization) $M$ for one or both states
becomes negative and hence unphysical.  Indeed, for the S state, $M<0$ for the entire 
$J_{3}=0.5$ curve, which is why we have not shown it in Fig.~\ref{E}(a). 

\begin{figure*}[!htb]
\mbox{
\subfigure[]{\scalebox{0.3}{\includegraphics[angle=270]{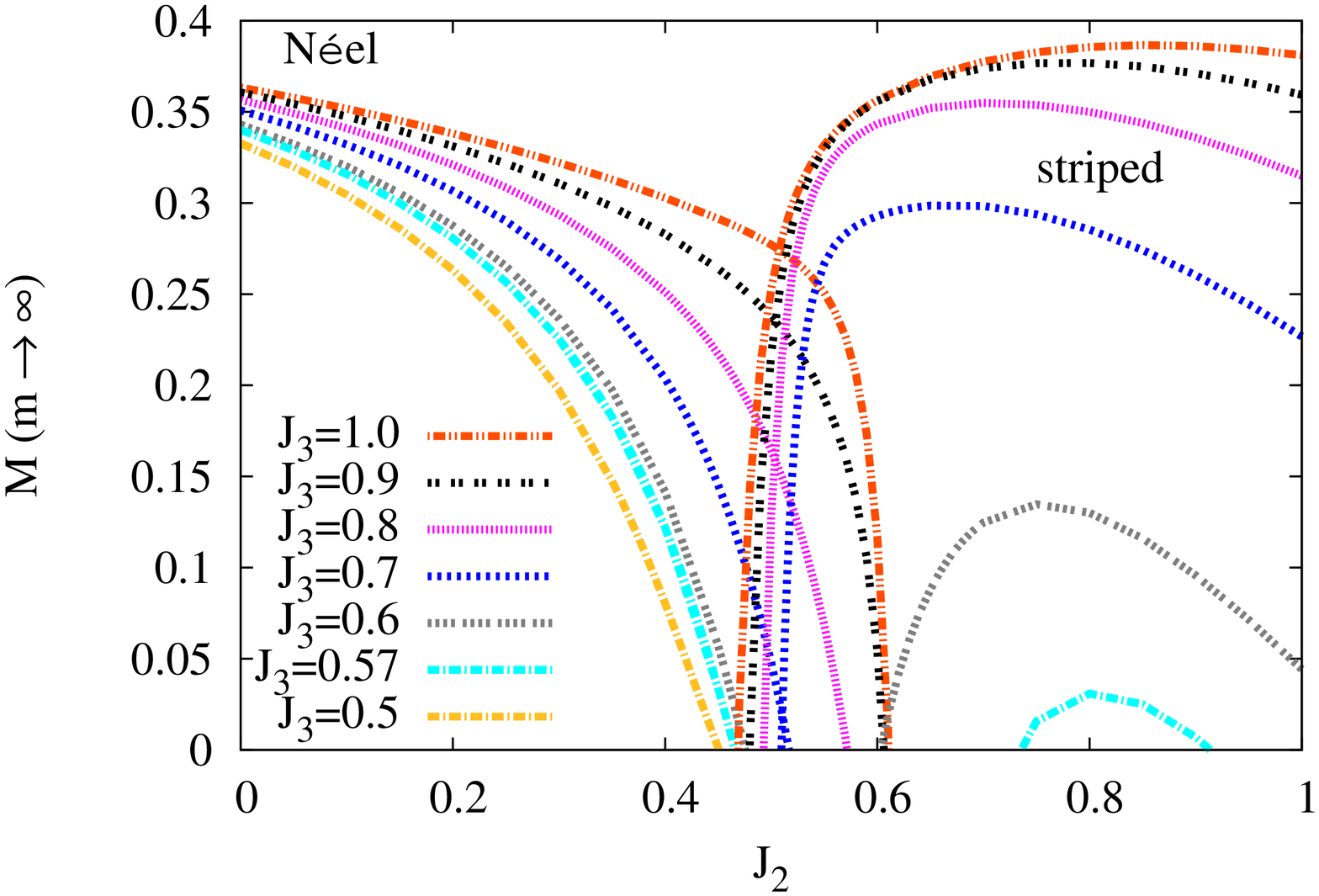}}}
\subfigure[]{\scalebox{0.3}{\includegraphics[angle=270]{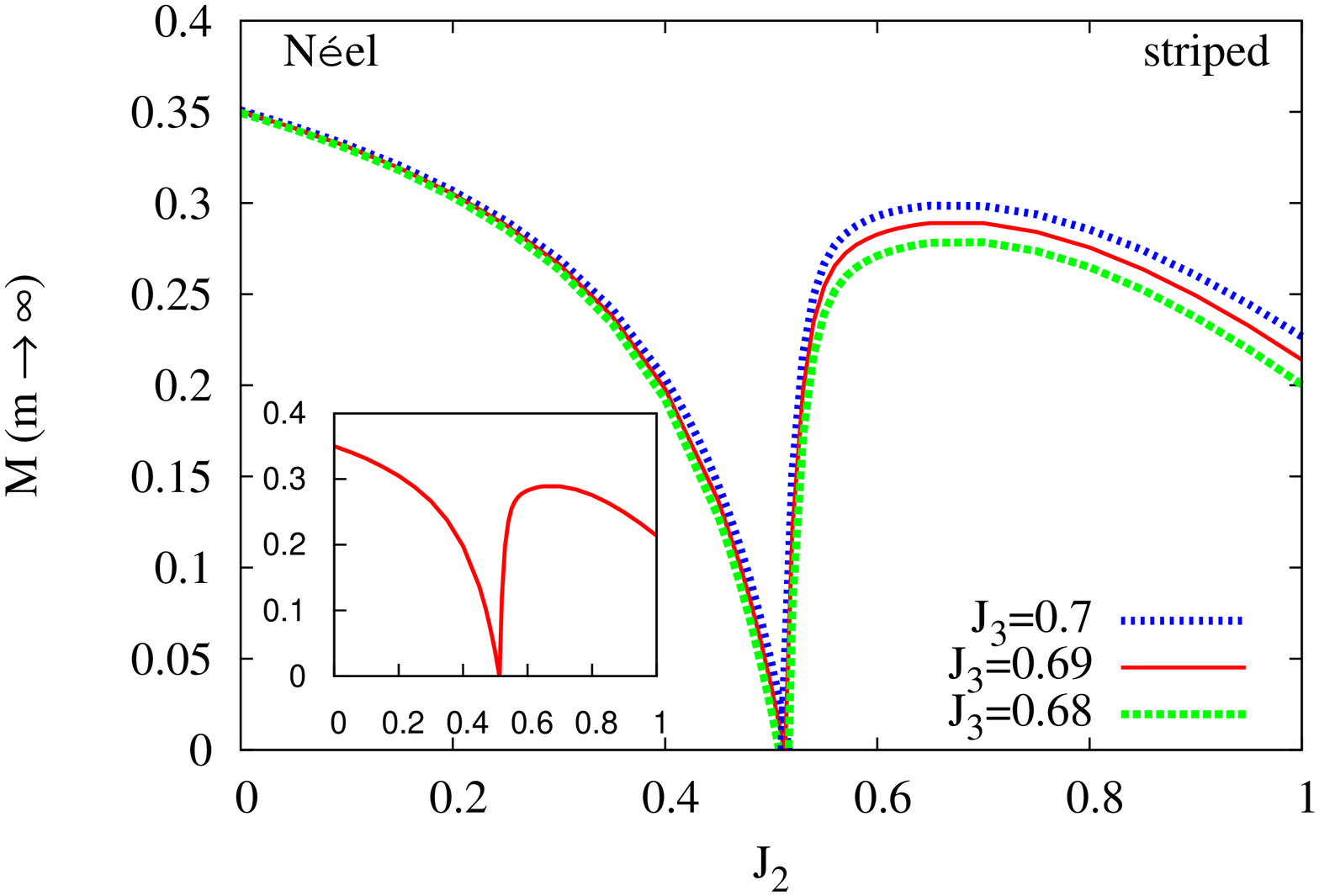}}}
}
\caption{ (Color online) Extrapolated CCM LSUB$\infty$ results for the
  GS magnetic order parameter, $M$, as a function of $J_{2}$, for
  various fixed values of $J_{3}$ in the range $0.5 \leq J_{3} \leq
  1.0$, for the N\'{e}el and the striped states of the
  spin-$\frac{1}{2}$ $J_{1}$--$J_{2}$--$J_{3}$ model on the honeycomb
  lattice (with $J_{1} \equiv 1$).  The extrapolated LSUB$m$ ($m
  \rightarrow \infty$) results are based on the extrapolation scheme
  of Eq.~(\ref{M_extrapo_frustrated}) and the calculated results with
  $m=\{6,8,10\}$.  
}
\label{M_Neel_striped}
\end{figure*}

In order to obtain more accurate values of the critical point we also show in
Figs.~\ref{M_Neel_striped}(a) and \ref{M_Neel_striped}(b) the
%%%%%%%%%%%%%%%%%%%%%%%%%%%%%%
%%%%%%%%%%%%%%%%%%%%%%%%%%%% 
curves for the extrapolated order parameters $M$ of the N and S states,
corresponding to values of $J_3$ shown in Figs.~\ref{E}(a) and \ref{E}(b) 
for the GS energy per spin, $E/N$.  The LSUB$\infty$ curves shown use
the LSUB$m$ results with $m=\{6,8,10\}$, together with the  extrapolation scheme of 
Eq.~(\ref{M_extrapo_frustrated}), which is appropriate for this highly frustrated regime.

We observe again that the curves intersect for values $J_{3} \gtrsim 0.69$, and that 
the corresponding values of $(J_{2},J_{3})$ are our best estimate for the phase boundary between
the N and S states, as shown on the phase diagram of Fig.~\ref{phase} by the points
denoted with (grey) convolution (eight-pointed star, \plusontimes) 
symbols.  For values
$J_{3} \lesssim 0.69$ the extrapolated order parameters of both the N and S phases become zero
before the curves intersect, revealing the presence of an intermediate phase in that regime.
The corresponding points in the case $J_{3} \lesssim 0.69$ where $M \to 0$ for the N and S phases
are shown in the phase diagram of Fig.~\ref{phase} by (red) times ($\times$) and (green) plus ($+$)
symbols respectively.
Our best value for the corresponding tricritical QCP comes from the data shown in 
Fig.~\ref{M_Neel_striped}(b), where it is seen to be at $(J_{2}^{c_1},J_{3}^{c_1})=(0.51 \pm 0.01,0.69 \pm 0.01)$, 
and where the error bars are estimates from a sensitivity analysis of the LSUB$m$ extrapolation scheme.

We note that the extrapolated order parameter $M$ becomes everywhere negative (i.e.,
for all values of $J_2$) for the S state for all values of $J_{3} \lesssim 0.55$, 
as may be seen from data similar to those shown in Fig.~\ref{M_Neel_striped}(a).
This is a clear first indication that the S state becomes unstable as the GS phase in this regime.
From a comparison with the corresponding classical model (i.e., in the limit $s \to \infty$) discussed in 
Sec.~\ref{model_section}, we might expect the S state to yield to the spiral state,
at least for sufficiently large values of $J_2$ in the present $s = \frac{1}{2}$ case.
We investigate this further in Sec.~\ref{results_spiral_striped} below.  It is clearly also expected
that the N\'{e}el (N) phase will not survive for large enough frustrating values of $J_{2} > 0$, and again from
a comparison with the classical model we expect that the spiral phase might exist in that case too.  Hence, 
we first make a comparison in Sec.~\ref{results_neel_spiral} of the N and spiral phases.

\subsection{N\'{e}el versus spiral phases}
\label{results_neel_spiral}
%%%%%%%%%%%%%%%%%%%%%%%%%%%%%%%%%%%%%%%%%%%%%%%%%%
\begin{figure*}
\mbox{
\subfigure[]{\scalebox{0.3}{\includegraphics[angle=270]{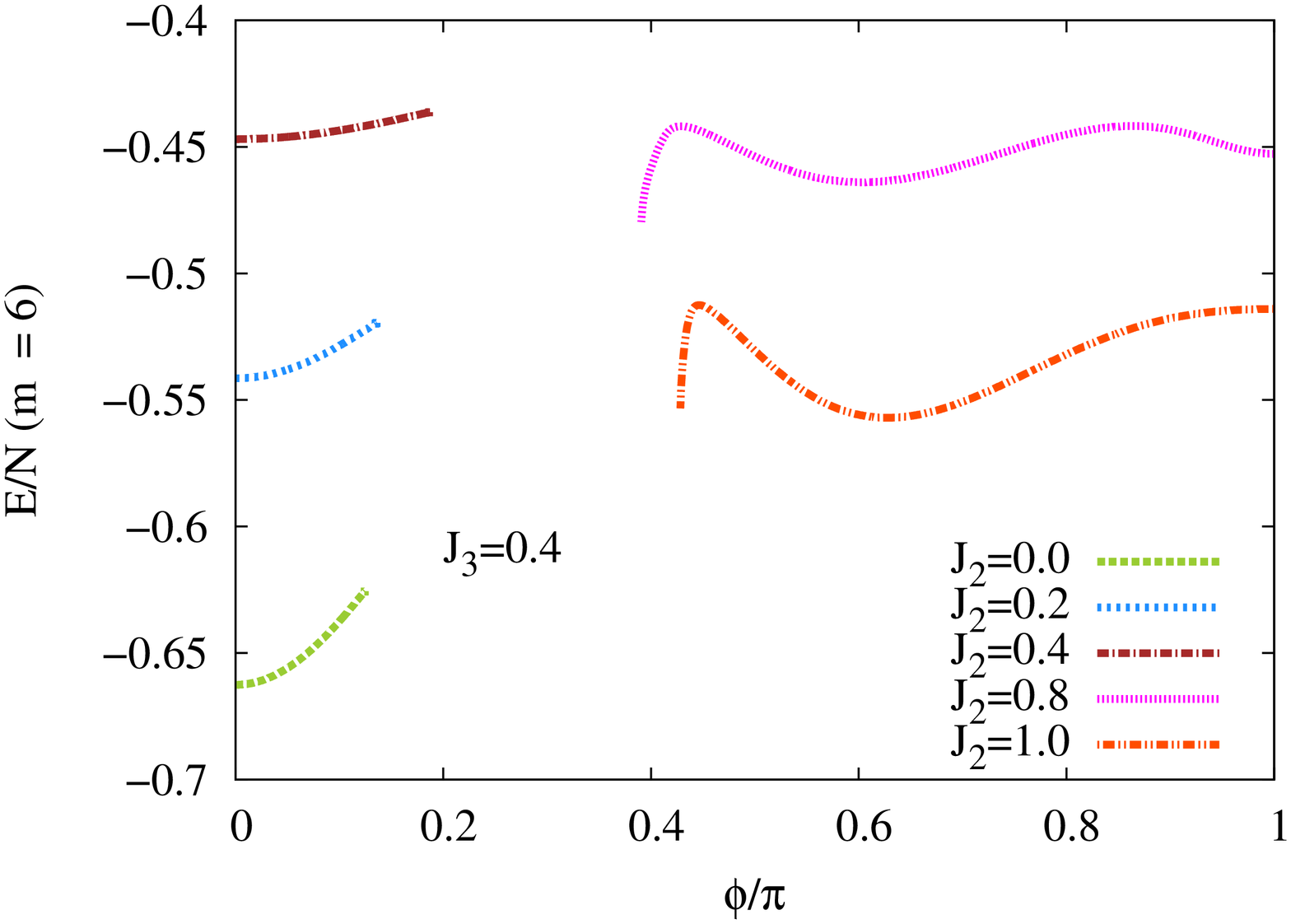}}}
\subfigure[]{\scalebox{0.3}{\includegraphics[angle=270]{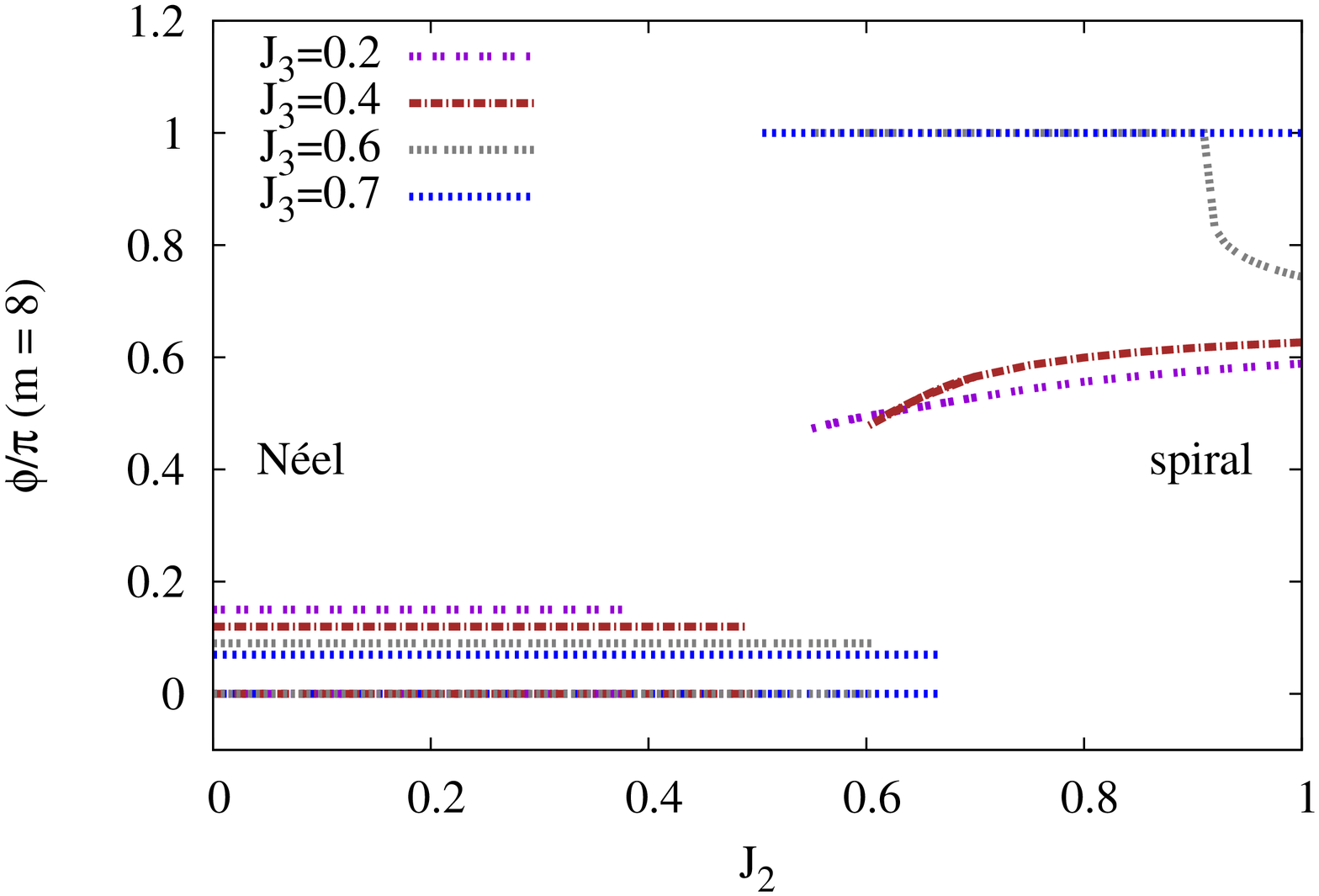}}}
}
\caption{
  (Color online) (a) LSUB6 results for the GS energy per spin of the spin-$\frac{1}{2}$ $J_{1}$--$J_{2}$--$J_{3}$ 
  model on the honeycomb lattice (with $J_{1} \equiv 1$), using the spiral state as CCM model state,
  as a function of the spiral pitch angle $\phi$, for some illustrative values
  of $J_{2}$ in the range $0 \leq J_{2} \leq 1$ and for a fixed value of $J_{3}=0.4$.  
  (b) The pitch angle $\phi = \phi_{{\rm LSUB}m}$ that minimizes the energy
  $E_{{\rm LSUB}m}(\phi)$ of the spiral state of the spin-$\frac{1}{2}$ 
  $J_{1}$--$J_{2}$--$J_{3}$ model on the honeycomb lattice (with $J_{1} \equiv 1$).  The CCM LSUB$m$
  results with $m=8$ are shown as functions of $J_2$ for several fixed values of $J_3$ in the range $0.2 \leq J_{3} \leq 0.7$.  The N\'{e}el results
  with $\phi=0$ are also shown artificially offset above the actual
  $\phi=0$ axis for the sake of clarity.
}
\label{E_vs_angle_Neel_vs_spiral}
\end{figure*}
%%%%%%%%%%%%%%%%%%%%%%%%%%%%%%%%%%%%%%%%%%
We start by analyzing the GS energy per spin, $E/N$, for the spiral state as a function of the
spiral pitch angle, $\phi$.  In our CCM calculations we choose the angle $\phi$, for each point in the
phase diagram where the spiral state exists, as the one that minimizes the energy estimate there.  Clearly
the minimizing angle in general also depends on the particular LSUB$m$ approximation being used with
the spiral state as CCM model state.  For example, we show in Fig.~\ref{E_vs_angle_Neel_vs_spiral}(a) the 
GS energy per spin, $E/N$, as a function of pitch angle $\phi$ in the LSUB6 approximation, for 
various illustrative values of $J_2$ at a fixed value of $J_{3}=0.4$ (and $J_{1} \equiv 1$).

We note first from Fig.~\ref{E_vs_angle_Neel_vs_spiral}(a) that for various fixed values of $J_{2}$ 
CCM solutions at a given LSUB$m$ level of approximation 
exist only for certain ranges of the spiral angle $\phi$.  For
example, for $J_{2}=0$ (and $J_{3}=0.4$), the CCM LSUB6 solutions based on the spiral
state exist only for $0 \leq \phi \lesssim 0.12\pi$.  In this
case, where the N\'{e}el state (i.e., where $\phi=0$) is the stable GS phase that minimizes the energy, if we try to force the system 
too far away from N\'{e}el collinearity the CCM equations themselves
become unstable in the sense that they no longer have a real solution.  We note too that
as $J_2$ is increased slowly (at fixed $J_{3}=0.4$), the minimum in the energy curve at $\phi=0$
becomes shallower, so that by the time $J_{2}=0.4$ it has almost disappeared.  This is a first
indication of the imminent instability of the N\'{e}el state as the GS phase if $J_2$ is increased slightly more.

Similarly, for $J_{2}=1$ (and $J_{3}=0.4$), the CCM LSUB6 solutions based on the spiral state exist only for 
$0.43 \lesssim \phi/\pi \leq 1$.  In this case, a spiral state (i.e., with a value $\phi \neq 0,\pi$) is 
the stable GS phase that minimizes the energy, and if we now try to force the system 
too close to the N\'{e}el regime, the CCM solution collapses.  We also observe
from Fig.~\ref{E_vs_angle_Neel_vs_spiral}(a) that for the smaller value $J_{2}=0.8$ (and $J_{3}=0.4$), while
the energy curve still shows a global minimum for a noncollinear spiral phase, it has now
also developed a secondary minimum at a value $\phi = \pi$ (i.e., that of the collinear striped state), which
indicates the proximity of the phase boundary between the spiral and striped states,
as we examine more fully in Sec.~\ref{results_spiral_striped} below.  

Conversely, as $J_3$ is increased further (for
fixed $J_2$), the spiral minimum becomes more pronounced, and as $J_{2} \rightarrow \infty$ the pitch angle
$\phi \rightarrow \frac{2}{3}\pi$.  This is as expected, since in this limit the model becomes two weakly connected
HAFMs on interpenetrating triangular lattices, with the classical ordering of NN spins oriented at angles
$\frac{2}{3} \pi$ with respect to one another on each sublattice.

From data such as that shown in Fig.~\ref{E_vs_angle_Neel_vs_spiral}(a) we can calculate in a given
LSUB$m$ approximation based on the spiral state as the CCM model state, the angle $\phi=\phi_{{\rm LSUB}m}$ that
minimizes the energy, $E_{{\rm LSUB}m}(\phi)$ for given values of the exchange coupling strengths.  For
example, in Fig.~\ref{E_vs_angle_Neel_vs_spiral}(b) we show the angle $\phi=\phi_{{\rm LSUB}8}$ from the
LSUB8 approximation, as a function of the parameter $J_2$ for several fixed values of the parameter $J_3$.
There is clear preliminary evidence that for values of $J_3$ below some upper critical value there is no
stable spiral solution for any value of $\phi \neq 0$ over a certain range of the parameter $J_2$, which itself depends on
$J_3$.  

Thus, we are led to expect a second tricritical QCP in the $(J_{2},J_{3}$) plane at $(J_{2}^{c_2},J_{3}^{c_2})$, with
$J_{3}^{c_2}<J_{3}^{c_1}$, such that: (a) for values $J_{3}>J_{3}^{c_1}$ the N and S states meet at a common phase boundary
discussed in Sec.~\ref{results_neel_striped} above, (b) for values $J_{3}^{c_1}>J_{3}>J_{3}^{c_2}$ there is an intermediate phase between
the N and S states, and (c) for values $J_{3}<J_{3}^{c_2}$ there is an intermediate phase between the N state and
the spiral state with $\phi \neq \pi$.  Thus, the QCP at $(J_{2}^{c_2},J_{3}^{c_2})$ is a tricritical point between
the N, S, and intermediate phases.  From the results discussed in Sec.~\ref{results_neel_striped} we now expect
that $J_{3}^{c_2} \approx 0.55$, and we discuss this further below.

\begin{figure}[!tb]
\includegraphics[width=6cm,angle=270]{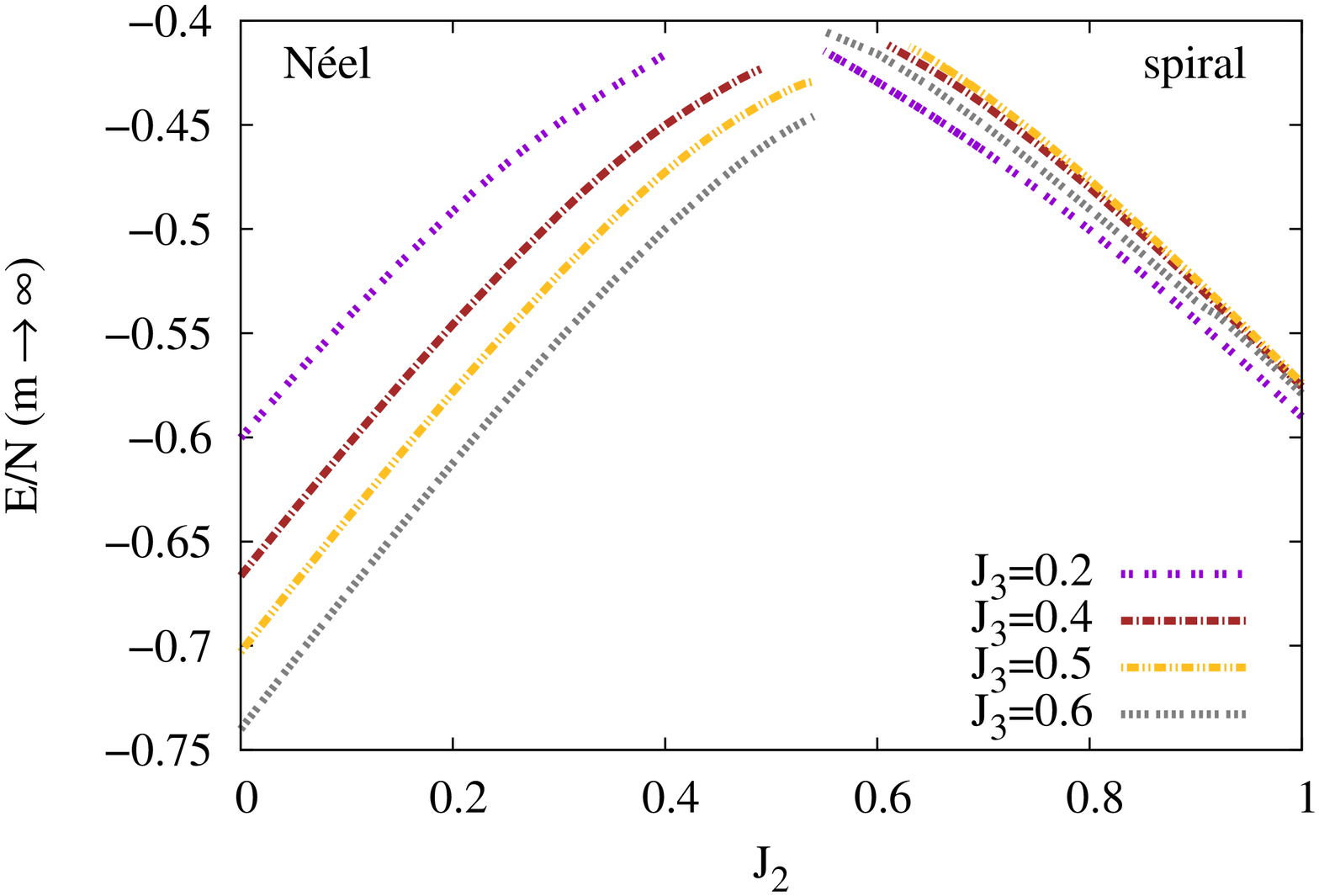}
\caption{ 
  (Color online) Extrapolated CCM LSUB$\infty$ results for the
  GS energy per spin, $E/N$, as a function of $J_{2}$, for various
  fixed values of $J_{3}$ in the range $0.2 \leq J_{3} \leq 0.6$, for
  the N\'{e}el and the spiral states of the spin-$\frac{1}{2}$
  $J_{1}$--$J_{2}$--$J_{3}$ model on the honeycomb lattice (with
  $J_{1} \equiv 1$).  The extrapolated LSUB$m$ ($m \rightarrow
  \infty$) results are based on the extrapolation scheme of
  Eq.~(\ref{E_extrapo}) and the calculated results with $m=\{6,8,10\}$
  for the N\'{e}el state, and with $m=\{4,6,8\}$ for the spiral state.
  For the spiral state the results use the pitch angle $\phi =
  \phi_{{\rm LSUB}m}$ that minimizes the energy $E=E_{{\rm LSUB}m}(\phi)$.  
}
\label{E_Neel_vs_spiral}
\end{figure}

Figure \ref{E_Neel_vs_spiral} shows our extrapolated CCM LSUB$\infty$
results for the GS energy per spin, $E/N$, as a function of $J_{2}$,
for various fixed values of $J_{3}$ in the range $0.2 \leq J_{3} \leq
0.6$, for the N\'{e}el and the spiral states.
%%%%%%%%%%%%%%%%%%%%%%%%%%%%%%%%%%%%%%%%
%%%%%%%%%%%%%%%%%%%%%%%%%%%%%%%%%%%%%%%%%%
Once again, the figure clearly illustrates the existence of an intermediate phase between the 
N\'{e}el and the spiral phases (including the striped state as a special case of the latter)
for values $J_{3}<J_{3}^{c_1}$.  

On a technical point we remark that for the spiral 
state the extrapolations are calculated using the LSUB$m$ calculated results with
$m=\{4,6,8\}$, rather than with the set $m=\{6,8,10\}$ used for the N\'{e}el state.  This is partly due
to the very high number, $N_{f} = 347287$, of configurations needed for the spiral state at the 
LSUB10 level of approximation, compared with the corresponding much smaller number, $N_{f} = 6237$, for the
N\'{e}el state, as seen from Table \ref{table_FundConfig}.  This difference is compounded by the fact
that for the spiral state we also need to do LSUB$m$ runs for each point in the phase space as a function
of the pitch angle $\phi$, in order to determine the angle $\phi = \phi_{{\rm LSUB}m}$ that minimizes the 
corresponding estimate for the energy, $E=E_{{\rm LSUB}m}(\phi)$.  This makes LSUB$m$ calculations for
the spiral state with $m \geq 10$ particularly demanding of computational resources.  

Nevertheless, we did perform LSUB10 calculations for the spiral state 
for the special case $J_{3}=0$ in our previous study
of the $J_1$--$J_2$ model,\cite{PHYLi:2012_honeyJ1-J2}
where we performed separate extrapolations using the LSUB$m$ results with $m=\{6,8,10\}$ and $m=\{4,6,8\}$.
We found that both extrapolations were in very good agreement with one another, and hence now feel
confident that the spiral-state extrapolations for the full $J_1$--$J_2$--$J_3$ model considered here with the limited set
$m=\{4,6,8\}$ will be equally robust, since it is now prohibitively expensive of computational resource to
perform LSUB10 calculations for the spiral state over the whole region of phase space where it is the stable 
GS phase.

We note that, as is usually the case, the CCM LSUB$m$ results for finite $m$ values
for a given phase extend beyond the actual physical LSUB$\infty$ boundary
for that phase.  Thus, the energy curves shown in Fig.~\ref{E_Neel_vs_spiral} for fixed values 
of $J_3$ terminate at certain values of $J_{2}$, which are determined by the termination
points of the highest LSUB$m$ approximations used in the extrapolations, beyond which no real solution exists for
the corresponding coupled CCM equations.  We note from Fig.~\ref{E_Neel_vs_spiral} that the maxima
in the energy curves occur close to these LSUB$m$ termination points for the largest $m$ values employed, which in turn
lie close to the physical (LSUB$\infty$) phase transition points.  

It has been suggested\cite{Albuquerque:2011}
that such an energy maximum approximately coincides with the avoided level crossing
to a different phase, in which case it could be taken as an approximation for the phase transition point of either
(the N\'{e}el or spiral) state to the intermediate (as yet unknown) state.  However, we do not
use this criterion here, since our results for the magnetic order parameter give us much
more accurate estimates, as we now discuss.
 
\begin{figure}[!tb]
\includegraphics[width=6cm,angle=270]{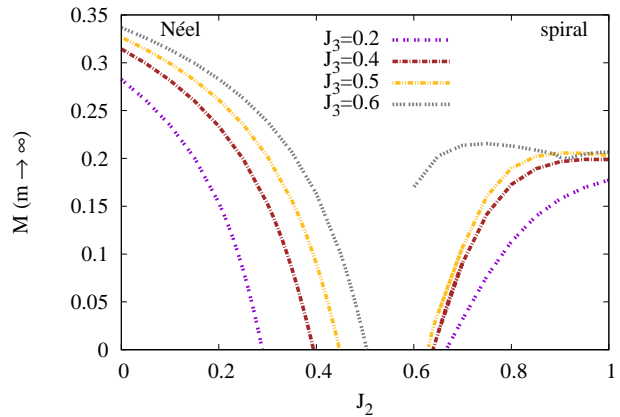}
\caption{ 
  (Color online) Extrapolated CCM LSUB$\infty$ results for the
  GS magnetic order parameter, $M$, as a function of $J_{2}$, for
  various fixed values of $J_{3}$ in the range $0.2 \leq J_{3} \leq
  0.6$, for the N\'{e}el and the spiral states of the
  spin-$\frac{1}{2}$ $J_{1}$--$J_{2}$--$J_{3}$ model on the honeycomb
  lattice (with $J_{1} \equiv 1$).  The extrapolated LSUB$m$ ($m
  \rightarrow \infty$) results are based on the extrapolation scheme
  of Eq.~(\ref{M_extrapo_frustrated_linearFit}) and the calculated
  results with $m=\{6,8,10\}$ for the N\'{e}el state, and with
  $m=\{4,6,8\}$ for the spiral state.  For the spiral state the
  results use the pitch angle $\phi = \phi_{{\rm LSUB}m}$ that
  minimizes the energy $E=E_{{\rm LSUB}m}(\phi)$.  
}
\label{M_Neel_vs_spiral}
\end{figure}

Thus, we show in Fig.~\ref{M_Neel_vs_spiral} our extrapolated CCM LSUB$\infty$ results for the GS
magnetic order parameters, $M$, for both the N\'{e}el and the spiral states,
as functions of $J_{2}$, for the same fixed values of $J_{3}$ shown in Fig.~\ref{E_Neel_vs_spiral} for the GS energy.
%%%%%%%%%%%%%%%%%%%%%%%%%%%%%%%%%%%
%%%%%%%%%%%%%%%%%%%%%%%%%%%%%%%%%%%%%%
The extrapolated LSUB$m$ ($m \rightarrow \infty$) results for the
N\'{e}el state are based on the extrapolation scheme of
Eq.~(\ref{M_extrapo_frustrated}) with $m=\{6,8,10\}$, whereas the
extrapolated results for the spiral state are based on the
extrapolation scheme of Eq.~(\ref{M_extrapo_frustrated_linearFit})
with $m=\{4,6,8\}$.  

As we indicated above, LSUB10 calculations are
prohibitively expensive for the spiral state, and hence we need
for extrapolation purposes to include the LSUB4 results.  When this
point is included, and the data set $m=\{4,6,8\}$ is thus employed, it
is clearly preferable to use the extrapolation scheme of
Eq.~(\ref{M_extrapo_frustrated_linearFit}) rather than that of
Eq.~(\ref{M_extrapo_frustrated}), so as not to give the $m=4$ result
too much weight.  However, in order to check our results, we have
performed LSUB10 calculations for the two values $J_{3}=0.2,0.4$.  For
these two values we have also made extrapolations using
Eq.~(\ref{M_extrapo_frustrated_linearFit}) with $m=\{6,8,10\}$, and
these are indicated in Fig.~\ref{phase} by the larger (orange) open
circles.  We find, very gratifyingly, that the two extrapolations
agree very well with one another at both values $J_{3}=0.2,0.4$, which
gives credence to our results using the data set $m=\{4,6,8\}$
elsewhere for the spiral state.

We note too that in our
earlier study\cite{PHYLi:2012_honeyJ1-J2} of this model with $J_{3}=0$
(and $J_{1} \equiv 1$), we also computed LSUB12 results for the
N\'{e}el state and found that the N\'{e}el order vanished at a value
$J_{2} \approx 0.207 \pm 0.003$ when we performed extrapolations
including the $m=12$ point.  It is this point that is shown in
Fig.~\ref{phase} by the larger (red) times ($\times$) symbol, although
the value obtained with the more limited data set $m=\{6,8,10\}$ used
for the results in Fig.~\ref{M_Neel_vs_spiral} is in good agreement
with it.  Similarly, in our earlier study of the model along the
$J_{3}=J_{2}$ line\cite{DJJF:2011_honeycomb} we also used LSUB$m$
results with $m=\{6,8,10,12\}$ to perform the extrapolations, and
found that in this case N\'{e}el order vanished at a value $J_{2}
\approx 0.466 \pm 0.005$, and this value is also shown in
Fig.~\ref{phase} by a larger (red) times ($\times$) symbol.

From curves such as those shown in Fig.~\ref{M_Neel_vs_spiral} we use the points where the extrapolated values of
the order parameter $M$ vanish, for various fixed values of $J_3$, to plot the phase boundaries of the N\'{e}el and
spiral phases denoted in Fig.~\ref{phase} by (red) times ($\times$) and (orange) 
open circle symbols respectively.  As expected from 
our previous discussion in Sec.~\ref{results_neel_striped}, a N\'{e}el-ordered phase exists for all values of $J_3$
up to some critical value of $J_2$ which marks its phase boundary.  For values $J_{3}<J_{3}^{c_1}\approx 0.69 \pm 0.01$
this phase borders a quantum paramagnetic phase, whereas for $J_{3}>J_{3}^{c_1}$ it borders the striped state at a first-order
phase transition boundary.  We also see from curves such as those shown in Fig.~\ref{M_Neel_vs_spiral} clear
evidence for an intervening phase between the N\'{e}el and spiral phases (with pitch angle $\phi \neq \pi$) everywhere
that the spiral phase exists.  Instead the spiral phase meets the striped phase along a common boundary (on
which $\phi=\pi$) for all values $J_{2} > J_{2}^{c_2} \gtrsim 0.65$.  There is thus a second tricritical point at
$(J_{2}^{c_2},J_{3}^{c_2})$, as we discuss more fully below in Sec.~\ref{results_spiral_striped}, at which the striped, spiral and 
quantum paramagnetic phases meet.
 
\subsection{Striped versus spiral phases}
\label{results_spiral_striped}
We first recall that classically (i.e., when $s \to \infty$) we have for this model (with $J_{1}\equiv 1$) that for 
a fixed value of $J_{2}>\frac{1}{2}$ the GS phase is the striped phase for $J_{3}>\frac{1}{2}J_{2}+\frac{1}{4}$ and
the spiral phase for $J_{3}<\frac{1}{2}J_{2}+\frac{1}{4}$.  There is a continuous phase transition between the 
two classical states along the boundary line, $J_{3}=\frac{1}{2}J_{2}+\frac{1}{4}\,, J_{2}\geq\frac{1}{2}$, on which 
the spiral pitch angle $\phi=\pi$.  Our results for the present $s=\frac{1}{2}$ model, as we shall see below, 
indicate that quantum fluctuations tend to stabilize the collinear order of the striped state to lower values of $J_3$, for fixed
$J_2$, than the classical limit.  Furthermore, as we shall see, the quantum fluctuations also seem to turn
the classical second-order transition into a quantum first-order one.

\begin{figure*}[!htb]
\mbox{
\subfigure[]{\scalebox{0.3}{\includegraphics[angle=270]{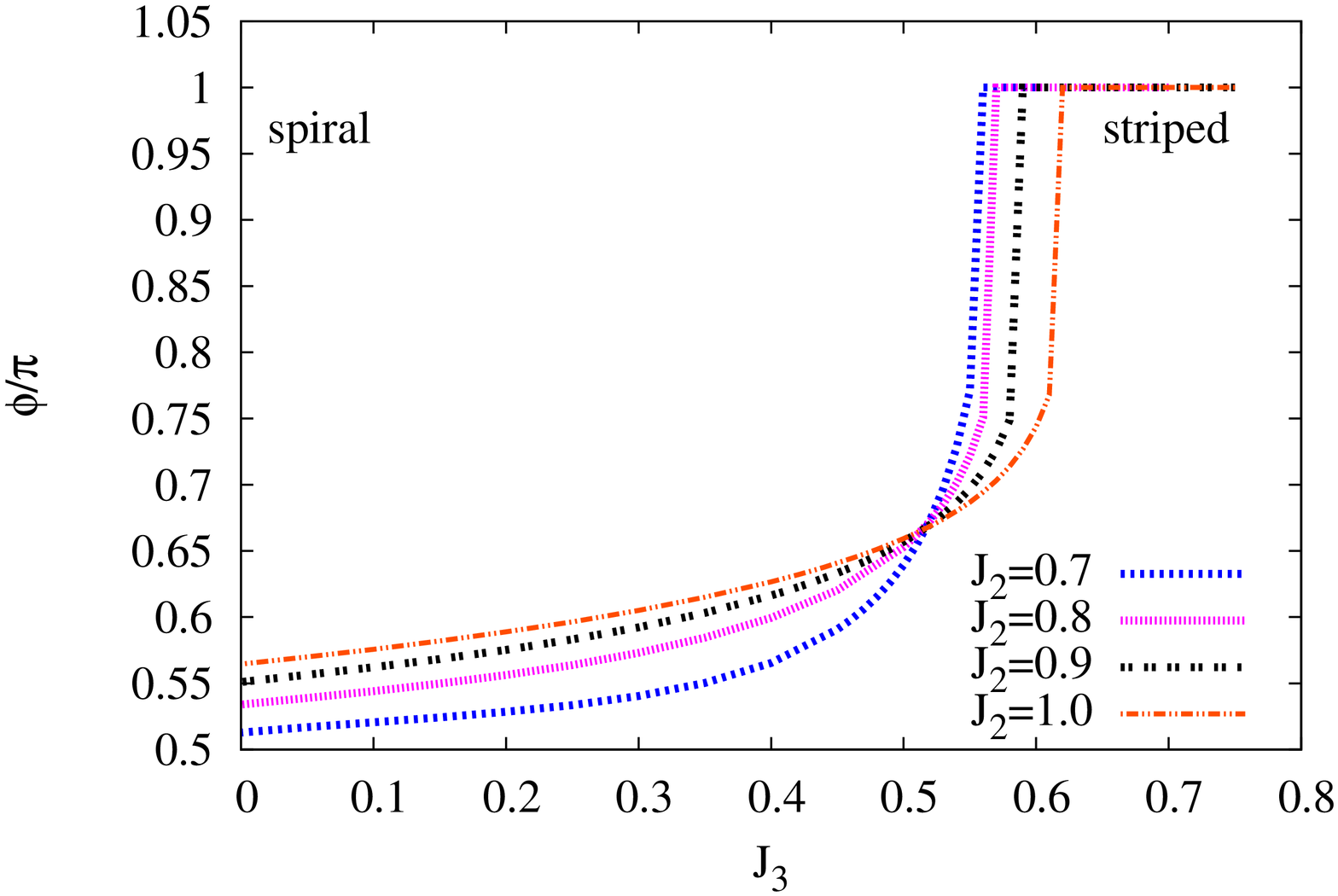}}}
\subfigure[]{\scalebox{0.3}{\includegraphics[angle=270]{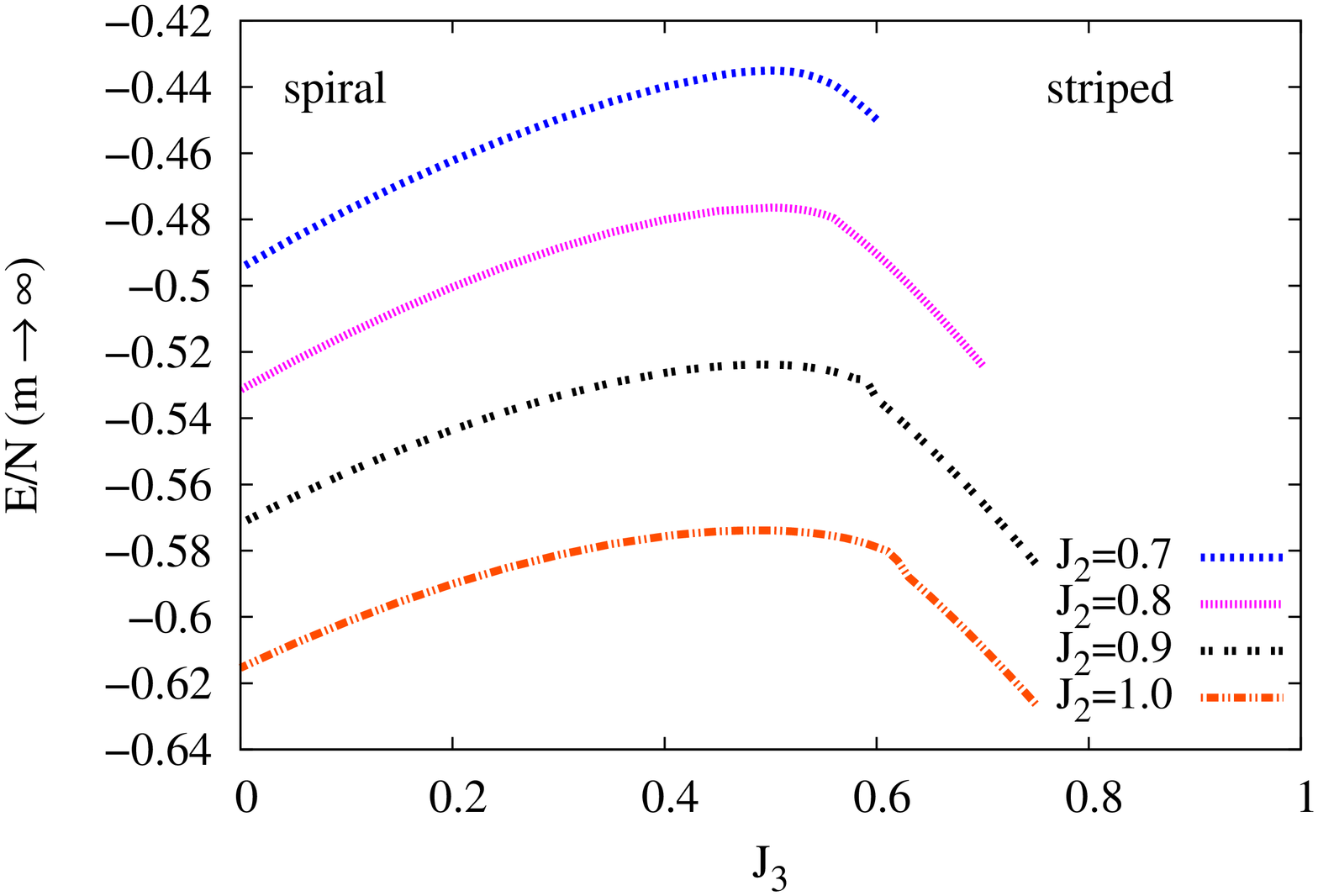}}}
}
\caption{
  (a) The angle $\phi = \phi_{{\rm LSUB}m}$ that minimizes the energy
  $E_{{\rm LSUB}m}(\phi)$ of the spin-$\frac{1}{2}$ 
  $J_{1}$--$J_{2}$--$J_{3}$ model on the honeycomb lattice (with $J_{1} \equiv 1$).  The CCM LSUB$m$
  results with $m=8$ are shown as functions of $J_3$ for several fixed values of $J_2$ in the range $0.7 \leq J_{3} \leq 1.0$.  
  Note that $\phi=\pi$ corresponds to the striped state.
  (b) Extrapolated CCM LSUB$\infty$ results for the
  GS energy per spin, $E/N$, as a function of $J_{3}$, for various
  fixed values of $J_{2}$ in the range $0.7 \leq J_{3} \leq 1.0$, for
  the spiral and the striped states of the spin-$\frac{1}{2}$
  $J_{1}$--$J_{2}$--$J_{3}$ model on the honeycomb lattice (with
  $J_{1} \equiv 1$).  The extrapolated LSUB$m$ ($m \rightarrow
  \infty$) results are based on the extrapolation scheme of
  Eq.~(\ref{E_extrapo}) and the calculated results with
  $m=\{4,6,8\}$. For the spiral state the
  results use the pitch angle $\phi = \phi_{{\rm LSUB}m}$ that
  minimizes the energy $E=E_{{\rm LSUB}m}(\phi)$.  
}
\label{E_spiral_striped}
\end{figure*}

Thus, we show in Fig.~\ref{E_spiral_striped}(a) the angle, $\phi=\phi_{{\rm LSUB8}}$ that
minimizes the energy, $E_{{\rm LSUB8}}(\phi)$, as a function of
$J_{3}$, using the spiral state as our CCM model state, for various fixed values of $J_2$.
%%%%%%%%%%%%%%%%%%%%%%%%%
%%%%%%%%%%%%%%%%%%%%%%%%%%%%%%%%%%%%
Very similar curves are found for other LSUB$m$ approximations.  We observe that, unlike in 
the classical case, where $\phi \to \pi$ continuously at the critical value, there is now a discontinuous
jump on the phase boundary.  Its origin lies in the double-minimum structure of the corresponding
energy curves (for fixed values of $J_2$ and $J_3$) as functions of the pitch angle $\phi$, comparable to that
shown in Fig.~\ref{E_vs_angle_Neel_vs_spiral}(a) for the case $J_{2}=0.8$, $J_{3}=0.4$.  

Clearly, if we consider the angle $\phi$ itself to be an order 
parameter (such that $\phi=\pi$ for striped order
and $\phi\neq 0,\pi$ for spiral order) the typical scenario for a first-order transition, as now seen here, is the 
emergence of such a two-minimum structure for $E/N$ as a function of $\phi$ for fixed coupling parameter strengths, one at
a value $\phi\neq\pi$ and the other precisely at $\phi=\pi$.
For a fixed value of $J_2$ we find, at a given LSUB$m$ level of approximation, that when $J_3$ is above a certain 
critical value the global minimum in the $E=E(\phi)$ curve is at $\phi=\pi$, whereas when $J_3$ is below this value
the global minimum is at the other minimum, $\phi\neq\pi$.

Figure~\ref{E_spiral_striped}(b) shows the corresponding extrapolated CCM LSUB$\infty$ results for the GS energy
per spin, $E/N$, of the spiral and striped states, as functions of the parameter $J_3$,  for the same fixed values of $J_2$ shown in 
Fig.~\ref{E_spiral_striped}(a).  The first-order transition between the spiral and striped states can clearly be seen to
occur close to, but not precisely at, the corresponding maxima in the energy curves.

\begin{figure}[!tb]
\includegraphics[angle=270,width=8cm]{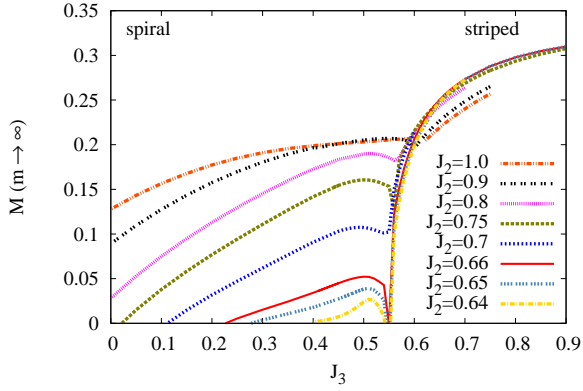}
\caption{ 
  (Color online) Extrapolated CCM LSUB$\infty$ results for the
  GS magnetic order parameter, $M$, as a function of $J_{3}$, for
  various fixed values of $J_{2}$ in the range $0.64 \leq J_{2} \leq
  1.0$, for the spiral and the striped states of the
  spin-$\frac{1}{2}$ $J_{1}$--$J_{2}$--$J_{3}$ model on the honeycomb
  lattice (with $J_{1} \equiv 1$).  The extrapolated LSUB$m$ ($m
  \rightarrow \infty$) results are based on the extrapolation scheme
  of Eq.~(\ref{M_extrapo_frustrated_linearFit}) and the calculated
  results with $m=\{4,6,8\}$.  For the spiral state the results use the pitch angle $\phi =
  \phi_{{\rm LSUB}m}$ that minimizes the energy $E=E_{{\rm LSUB}m}(\phi)$.  
}
\label{M_spiral_vs_Striped}
\end{figure}

As before, the actual phase boundary is most clearly seen from our similarly extrapolated CCM LSUB$\infty$ 
results for the magnetic order parameter, $M$, which are shown in Fig.~\ref{M_spiral_vs_Striped}. 
%%%%%%%%%%%%%%%%%%%%%%%
%%%%%%%%%%%%%%%%%%%%%%
We note that for all values of $J_{2} \gtrsim 0.66$ there is a clear and sharp minimum in
the magnetic order parameter at the phase transition point in the parameter $J_3$ where the striped and spiral phases
meet.  These points are indicated by the (cyan) open triangle ($\triangle$)
symbols in the phase diagram shown in Fig.~\ref{phase}.

At the value $J_{2} \approx 0.66$ the two curves meet at $M=0$.  We note that
for this value of $J_2$ the magnetic order parameter $M$ for the spiral state is very small (and positive) for all values of 
$J_3$, and that as $J_2$ is decreased further the spiral state rapidly disappears altogether for $J_{2} \lesssim 0.635$.
In the very narrow regime $0.635 \lesssim J_{2} \lesssim 0.66$, we see from Fig.~\ref{M_spiral_vs_Striped}
that there appears to be an intrusion of the intermediate (quantum paramagnetic) phase, as shown in Fig.~\ref{phase} by the appearance of both
(cyan) open triangle ($\triangle$) and (orange) open circle ($\bigcirc$) symbols 
at the striped-spiral phase boundary at the two values $J_{2}=0.64,0.65$.  It seems almost sure,
however, that this effect arises from our extrapolations, and is an indication of the (small) errors inherent in them.
Our best estimate from the results shown in Fig.~\ref{M_spiral_vs_Striped} is thus that the second tricritical QCP,
where the spiral, striped and quantum paramagnetic phases meet, occurs at $(J_{2}^{c_2},J_{3}^{c_2})=(0.65 \pm 0.02,0.55 \pm 0.01)$.

We also note from Fig.~\ref{M_spiral_vs_Striped} that for values $0.635 \lesssim J_{2} \lesssim 0.77$ and $J_{3}>0$ the
magnetic order parameter $M$ of the striped state becomes zero at a lower critical value of $J_3$.  These lower values
in each case are shown in Fig.~\ref{phase} by the same (orange) open circle ($\bigcirc$)
symbols as we discussed previously in Sec.~\ref{results_neel_spiral}.
We note that for the special case $J_{3}=0$ that we investigated earlier,\cite{PHYLi:2012_honeyJ1-J2} the spiral state is actually 
unstable, since the anti-N\'{e}el state was seen to have lower energy for all values of $J_2$ in the range investigated, namely
$J_{2} \leq 1$, where solutions for the spiral state could be found.  From continuity, we expect that the
anti-N\'{e}el state should remain the stable GS phase for small enough values of $J_3$ below some critical
value for each fixed value of $J_2$, above which value the spiral phase then becomes the stable GS phase.  
Thus we are led to expect that there might exist a third tricritical QCP at $(J_{2}^{c_3},J_{3}^{c_3})$ between
the spiral, quantum paramagnetic, and anti-N\'{e}el GS phases.  
We examine this further in Sec.~\ref{results_spiral_aN} below.

\subsection{Spiral versus anti-N\'{e}el phases}
\label{results_spiral_aN} 
\begin{figure}[!tb]
\includegraphics[angle=270,width=8cm]{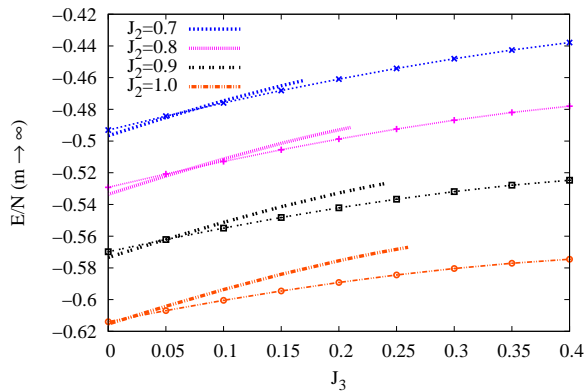}
\caption{
  (Color online) Extrapolated CCM LSUB$\infty$ results for the
  GS energy per spin, $E/N$, as a function of $J_{3}$, for various
  fixed values of $J_{2}$ in the range $0.7 \leq J_{2} \leq 1.0$, for
  the anti-N\'{e}el and the spiral states of the spin-$\frac{1}{2}$
  $J_{1}$--$J_{2}$--$J_{3}$ model on the honeycomb lattice (with
  $J_{1} \equiv 1$).  The extrapolated LSUB$m$ ($m \rightarrow
  \infty$) results are based on the extrapolation scheme of
  Eq.~(\ref{E_extrapo_linearFit}), and the calculated results with
  $m=\{4,6,8\}$ in both cases.  For the spiral state the results use the pitch angle $\phi =
  \phi_{{\rm LSUB}m}$ that minimizes the energy $E=E_{{\rm LSUB}m}(\phi)$.  
  We note that in all cases curves without symbols attached refer to the anti-N\'{e}el state,
  whereas the corresponding curves with symbols refer to the spiral state.
}
\label{E_cross_spiral_vs_aN}
\end{figure}

In Fig.~\ref{E_cross_spiral_vs_aN} we show the extrapolated CCM
LSUB$\infty$ results for the GS energy per spin, $E/N$, of both the spiral and anti-N\'{e}el states,
as functions of the parameter $J_{3}$,  for various fixed values of the parameter $J_2$ in the range
$0.7 \leq J_{2} \leq 1.0$. 
%%%%%%%%%%%%%%%%%%%%%%%%%%%%%%%%%
%%%%%%%%%%%%%%%%%%%%%%%%%%%%%%%%%
Although the energy differences are small for each fixed value of $J_2$, the results at each LSUB$m$ level,
as well as the extrapolated results, clearly show an energy crossing point.  These 
energy crossing points are thus our first estimates of the phase boundary points between the 
spiral and anti-N\'{e}el states.

On a technical point, we have noted previously that CCM LSUB$m$
calculations for the spiral state are computationally very expensive
for values of the truncation index $m \geq 10$.  Thus, we are for the
most part restricted to the data set $m=\{4,6,8\}$ for the spiral
state, although we have performed a very few calculations for a few
select values in the parameter space with $m=10$.  With only three
data points to fit to an extrapolation formula, a two-term
extrapolation fit (such as those in Eqs.~(\ref{E_extrapo_linearFit})
and (\ref{M_extrapo_frustrated_linearFit}), for example) can often be
preferable in practice to a three-term fit (such as their counterparts
in Eqs.~(\ref{E_extrapo}) and (\ref{M_extrapo_frustrated}),
respectively).  This is particularly the case when one of the data
points is either far from the limiting case or when it does not
represent all of the features of the system as well as the remaining,
more accurate points, as is the case here for the $m=4$ points.  

Thus, since the energy differences of the spiral and aN states are
relatively small, we have found it preferable to use the extrapolation
scheme of Eq.~(\ref{E_extrapo_linearFit}) in this case, and to employ
the same data set with $m=\{4,6,8\}$ for both (aN and spiral) phases,
even though results with $m=10$ are more readily available for the aN
state.  We have, however, demonstrated that the results so obtained
are robust and reliable, by making further checks in some limited test
cases using the extrapolation scheme of Eq.~(\ref{E_extrapo}) fitted
to the results $m=\{4,6,8,10\}$ or the extrapolation scheme of
Eq.~(\ref{E_extrapo_linearFit}) fitted to the results $m=\{6,8,10\}$,
for example.

We note that real CCM LSUB$m$ solutions based on the aN state as model state cease to exist, for a fixed value of
$J_2$, above some termination value in the parameter $J_3$ that itself depends on the truncation index $m$, 
just as we have indicated above for other phases.  These LSUB8 terminations are what cause our extrapolations
for the aN state to be shown only up to certain values of $J_3$ for each curve shown in 
Fig.~\ref{E_cross_spiral_vs_aN}.  In each case, the LSUB$m$ solution with a finite value of $m$ 
extends further into the region where the aN solution actually ceases to exist (i.e., to after the energy
crossing point with the spiral phase).  Presumably, in the $m \to \infty$ limit, the LSUB$m$ termination
points for the aN phase would coincide with the phase boundary with the spiral phase.  Simple heuristic extrapolations 
based on the results with $m=\{4,6,8\}$ agree well with the energy crossing points, and are hence
entirely consistent with this hypothesis.

\begin{figure}[!tb]
\includegraphics[angle=270,width=9cm]{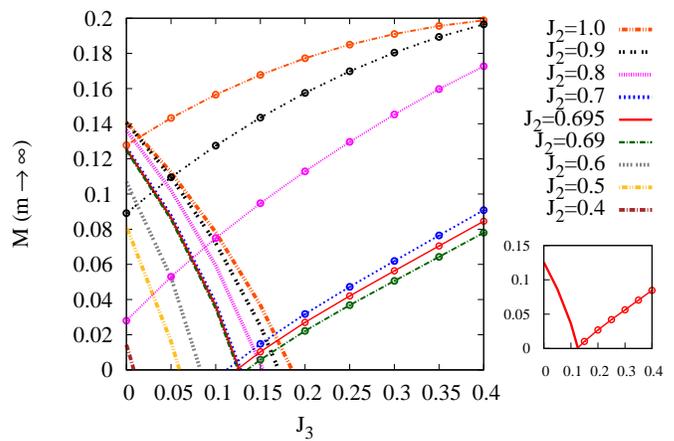}
\caption{
  (Color online) Extrapolated CCM LSUB$\infty$ results for the
  GS magnetic order parameter, $M$, as a function of $J_{3}$, for various
  fixed values of $J_{2}$ in the range $0.4 \leq J_{3} \leq 1.0$, for
  the anti-N\'{e}el and the spiral states of the spin-$\frac{1}{2}$
  $J_{1}$--$J_{2}$--$J_{3}$ model on the honeycomb lattice (with
  $J_{1} \equiv 1$).  The extrapolated LSUB$m$ ($m \rightarrow
  \infty$) results are based on the extrapolation scheme of
  Eq.~(\ref{M_extrapo_frustrated_linearFit}), and the calculated results with
  $m=\{4,6,8\}$ for both the aN and spiral phases for values $J_{2}=0.69,0.695,0.7,0.8,0.9,1.0$; 
  and with $m=\{6,8,10\}$ for the aN phase for values $J_{2}=0.4,0.5,0.6$.  For the spiral 
  state the results use the pitch angle $\phi =
  \phi_{{\rm LSUB}m}$ that minimizes the energy $E=E_{{\rm LSUB}m}(\phi)$.  
  We note that in all cases curves without symbols attached refer to the anti-N\'{e}el state,
  whereas the corresponding curves with symbols refer to the spiral state.}
\label{M_spiral_vs_aN}
\end{figure}

We note from Fig.~\ref{E_cross_spiral_vs_aN} that as the value of the parameter $J_2$ is decreased
towards the lower limiting value $J_{2} \approx 0.635$, below which the spiral state ceases to
exist, the energy curves for the aN and spiral phases lie increasingly close to one another,
and hence the position of the crossing point becomes increasingly difficult to
determine with high precision.  Accordingly, we expect that a better indicator
of the phase boundary might be obtained from a comparison of the magnetic order parameters
of the two states, as now shown in Fig.~\ref{M_spiral_vs_aN}.
%%%%%%%%%%%%%%%%%%%%%%%%%%%
%%%%%%%%%%%%%%%%%%%%%%%%%%%

We see very clearly that for values of $J_{2} \gtrsim 0.69$ the curves for the magnetic
order parameters $M$ of the aN and spiral phases cross in the physical regime (i.e., at a positive value of $M$).
It is these crossing points that are our best estimates for the corresponding points on 
the boundary between the two phases, and these are shown in the phase diagram of Fig.~\ref{phase} by 
(magenta) times ($\times$) symbols.  For values $J_{2} \gtrsim 0.8$, these values are in 
excellent quantitative agreement with the corresponding energy crossing points from
Fig.~\ref{E_cross_spiral_vs_aN}.  For smaller values of $J_2$, down to the value
$J_{2} \approx 0.635$ below which the spiral state ceases to exist for any value of $J_3$,
the energy crossing points become increasingly difficult to estimate accurately,
as discussed above, and generally lie slightly below the much more accurate values
obtained from Fig.~\ref{M_spiral_vs_aN}, although they are still in good qualitative agreement 
with them.  

Our best estimate for the position of the third tricritical QCP, 
$(J_{2}^{c_3},J_{3}^{c_3}) = (0.69 \pm 0.01, 0.12 \pm 0.02)$, which marks the point where the spiral and aN phases meet 
the QP phase, comes from curves such as those shown in Fig.~\ref{M_spiral_vs_aN}.

For values $J_{2} < J_{2}^{c_3} \approx 0.69$, we use the corresponding values of $J_3$ at which
the magnetic order parameter $M \to 0$ for the aN phase, as shown in Fig.~\ref{M_spiral_vs_aN}, to
find the phase boundary between the aN and QP phases.  The corresponding points are
shown in the phase diagram of Fig.~\ref{phase} by (blue) plus ($+$) symbols.

\subsection{The quantum paramagnetic (PVBC?) phase(s)}
\label{results_paramagnet}
From the results presented so far we have seen that in the parameter space window
$J_{2},J_{3} \in [0,1]$ the spin-$\frac{1}{2}$ $J_{1}$--$J_{2}$--$J_{3}$ HAFM
on the honeycomb lattice with $J_{1} \equiv 1$ has regions of five different GS phases.  Four of these 
(viz., the N, S, aN, and spiral phases) are quasiclassical in nature, and they almost completely surround the fifth
QP phase, as shown in Fig.~\ref{phase}, with each of them sharing a boundary with the (almost) enclosed 
region of the QP phase.  (Indeed, it seems likely that if the diagram were extended
slightly to negative values of $J_3$, the QP region would be seen to be entirely enclosed.)  In the
window $J_{2},J_{3} \in [0,1]$ there are three tricritical QCPs (and it seems likely that a fourth, which 
marks the meeting of the N, aN and QP phases, will occur just outside the window).  

From our current results discussed here the question still 
remains open, however, as to the exact nature of this phase.  What we know from previous 
work\cite{DJJF:2011_honeycomb,PHYLi:2012_honeyJ1-J2} that employed the same CCM techniques as here is that
the QP phase appears to be have PVBC ordering at least at four points along its boundary.  These include the
two points marked with the larger (red) times ($\times$) symbols in Fig.~\ref{phase} along its boundary with the
N state where it crosses both the $J_{3}=0$ axis and the $J_{3}=J_{2}$ line, the point marked with the larger (green) 
plus ($+$) symbol along its boundary with the S state where it crosses the $J_{3}=J_{2}$ line, and
the point marked with the larger (blue) plus ($+$) symbol along its boundary with the aN
state where it crosses the $J_{3}=0$ axis.  

Those points were identified as lying on a phase boundary with the PVBC state
by calculating, within the same CCM LSUB$m$ approximations as used to calculate the phase transition points that marked
the vanishing of the magnetic order parameter $M$ in each case (i.e., for the N, S, and aN phases respectively), the 
susceptibility of the respective phases against the formation of PVBC order.  We showed, within the accuracy of our results,
that each of the above four points where the respective magnetic order parameter 
of each quasiclassical phase goes to zero coincide with
the points at which the corresponding susceptibility of the state to PVBC order becomes infinite.  

In principle we could now repeat those calculations for the PVBC susceptibility parameter for all points
along the phase boundary of the QP state with the four quasiclassical states.  However, that would
be particularly costly of computing resources for the spiral state.  Even if we were to do so and hence show
that the entire boundary region has PVBC order, we
could still not be sure that the QP region was entirely PVBC-ordered, since 
there might still exists regions of other phases with other forms of order, possibly even of an
exotic (spin-liquid) variety.  The full characterization of the ordering within the QP phase(s)
bounded by the four phases with quasiclassical ordering is an extremely challenging problem, and 
one that is essentially outside the scope of the present investigation.

\section{SUMMARY AND DISCUSSION}
\label{summary}
In this paper we have studied the spin-$\frac{1}{2}$ HAFM on the honeycomb lattice with NN, NNN, and NNNN exchange
interactions, namely the so-called $J_{1}$--$J_{2}$--$J_{3}$ model described by the 
Hamiltonian of Eq.~(\ref{eq1}).  We have investigated the full phase 
diagram of the model, in the case where all the bonds are antiferromagnetic in nature (i.e., $J_{n}>0,\, n=1,2,3$),
based on a combination of CCM techniques described in detail in Secs.~\ref{CCM} and \ref{results}. 
In particular we have set $J_{1} \equiv 1$ to set the overall energy scale, and we have restricted attention here
to the window $J_{2},J_{3} \in [0,1]$ for the remaining parameters.  

Our results are summarized in the phase diagram of Fig.~\ref{phase}.  In the window $J_{2},J_{3} \in [0,1]$ 
we find the five stable GS phases shown.  Four of them are quasiclassical in nature, in the 
sense that they have counterparts in the classical version ($s \to \infty$)
of the model.  These comprise three
phases showing collinear AFM order, viz., the N\'{e}el (N), striped (S), and anti-N\'{e}el states
depicted in Figs.~\ref{model}(a), (b), and (d) respectively, plus a noncollinear spiral phase depicted in
Fig.~\ref{model}(c).  In the classical version of the model, however, only the three phases with N, S, and spiral
order exist in the examined window $J_{2},J_{3} \in [0,1]$.  In the classical model the phase with aN 
order exists only for a part of the phase space where $J_{3}<0$.  We find that quantum fluctuations
tend to stabilize this collinear aN phase at the expense of the spiral phase, in keeping
with the very general observation that quantum fluctuations always seem to favor collinear
phases over noncollinear ones.  The fifth phase shown in Fig.~\ref{phase} is a magnetically disordered, 
or quantum paramagnetic (QP), phase that has no classical counterpart.  

The QP phase has phase boundaries with
each of the four quasiclassical phases, with three tricritical QCPs occurring in the window
$J_{2},J_{3} \in [0,1]$, and a fourth presumably occurring just outside the window with a small
negative value of $J_3$.  The boundaries of the QP phase with each of the S, aN, and spiral phases
appear to delimit first-order transitions, whereas there are strong indications that the boundary of 
the QP phase with the N phase delimits a continuous phase transition.  At two points along this boundary,
viz., along the lines $J_{3}=0$ and $J_{3}=J_{2}$, there are strong indications that the QP phase there
has PVBC order.  Since the N and PVBC phases break different symmetries we have argued that the transitions there favor
the deconfinement scenario.  Nevertheless, we cannot entirely exclude even at these points the possibilities
of a very weak first-order transition or that the transition proceeds through a very narrow region of
intervening phase (possibly even of an exotic spin-liquid variety).  

We have found that all of the three remaining phase transition lines in the $J_{2},J_{3} \in [0,1]$ window
between the pairs of quasiclassical states shown in Fig.~\ref{phase} (viz., between the N and S, S and spiral,
and the spiral and aN phases) are first-order in nature.  Rather strikingly, quantum effects in the
$s=\frac{1}{2}$ model turn the continuous transition between the spiral and S states of the classical
($s \to \infty$) model into a first-order one.  

Whereas in the classical model the only three
phases present in the $J_{2},J_{3} \in [0,1]$ window (viz., the N, S, and spiral phases) meet at a single
tricritical point at $(J_2^{c,{\rm cl}}, J_3^{c,{\rm cl}})=(0.5,0.5)$, there are now three
tricritical QCPs in the same window for the $s=\frac{1}{2}$ model.  The classical tricritical point
separates into two tricritical QCPs at $(J_{2}^{c_1},J_{3}^{c_1})=(0.51 \pm 0.01,0.69 \pm 0.01)$ between
the N, S, and QP phases, and at $(J_{2}^{c_2},J_{3}^{c_2})=(0.65 \pm 0.02,0.55 \pm 0.01)$
between the S, spiral, and QP phases.  A third tricritical QCP at 
$(J_{2}^{c_3},J_{3}^{c_3}) = (0.69 \pm 0.01, 0.12 \pm 0.02)$, is identified for
the spin-$\frac{1}{2}$ model between the spiral, aN, and QP phases.

In overall terms our results for the phase diagram are in good agreement with other very recent
studies of this model. For example, a study using a combination of various exact diagonalization (ED) and
self-consistent cluster mean-field (SCCMF) techniques\cite{Albuquerque:2011} finds a phase diagram with basically the
same five phases as we identify in the window $J_{2},J_{3} \in [0,1]$.  There is good agreement with
the positions of the two tricritical QCPs at $(J_{2}^{c_1},J_{3}^{c_1})$ and
$(J_{2}^{c_2},J_{3}^{c_2})$.  The main difference seems to be in the position of the third
tricritical point at $(J_{2}^{c_3},J_{3}^{c_3})$.  Although both sets of calculations seem
to be in good agreement about the boundary of the aN phase, the ED results generally place the
boundary between the spiral and QP phases to lower values of $J_2$ such that the spiral phase
occupies a larger region of phase space than our own calculations indicate.  We note, however,
that ED calculations are especially difficult for noncollinear phases, since the finite
lattices used do not so easily sample such noncollinear phases.

A further recent study of the model, using an unbiased pseudo-fermion functional 
renormalization group (PFFRG) method,\cite{Reuther:2011} gives a phase diagram again in good overall 
agreement with ours, and with a phase boundary between the spiral and QP phases 
now in closer agreement with ours than from the ED results.\cite{Albuquerque:2011}
Again, there is also good agreement with the phase boundaries involving the N and S states,
and the positions of the two tricritical QCPs at $(J_{2}^{c_1},J_{3}^{c_1})$ and
$(J_{2}^{c_2},J_{3}^{c_2})$.  The only qualitative disagreement is that the PFFRG study finds
no evidence in the $J_{2},J_{3} \in [0,1]$ window for the aN phase.  Nevertheless, this 
study did find that for larger values of $J_2$ and small values of $J_3$ there were
large incommensurability shifts from the spiral phase and, furthermore, that
there was evidence for $J_{2} \gtrsim 0.4$ of sizeable staggered dimer order.  Such
staggered dimer order is often difficult in practice to distinguish from the
striped AFM order.

Series expansion (SE) techniques have also been applied to this model recently.\cite{Oitmaa:2011}
In particular, expansions were performed around our N, S, and spiral states, as well as about
the second classical spiral phase and the staggered dimer 
valence-bond crystal (SDVBC) state, which is also called the lattice nematic
state.  The results from the SE analysis are much more qualitative than ours or those of the
ED+SCCMF\cite{Albuquerque:2011} and PFFRG\cite{Reuther:2011} studies.  Nevertheless, the SE
study is also in broad agreement, with the exception again of finding no evidence for the
aN state (which might have been located for the only value $J_{3}=0$ that was studied by those authors
with the second spiral phase that exists classically in the range $\frac{1}{6}<J_{2}<\frac{1}{2}$ for
$J_{3}=0$, and for which the aN phase is a limiting collinear form as discussed in 
Sec.~\ref{model_section}).  

It is clear that the existence or not of the aN phase in the $J_{2},J_{3} \in [0,1]$ window 
is one of the points on which there is still disagreement between various studies.  It seems
clear that, if it does indeed exist, as we argue here, it becomes unstable at very small
values of $J_3$ for all values of $J_{2} < 1$ for which it exists.  We note that a quite different 
recent ED study\cite{Mosadeq:2011} of the model (along the $J_{3}=0$ line only), which
used a NN singlet valence-bond basis, found very similar
critical points to ours for the boundaries of the QP phase, but found that while the QP phase
was bounded on one side (for smaller values of $J_2$) by the N phase, it was bounded on the other
side by a SDVBC phase.  However, these results would again be equally consistent
with the identification of this phase as our aN phase, since their interpretation is
strongly biased by their choice of basis.  

A quite separate study of the case $J_{3}=0$ has 
also been performed recently based on an entangled-plaquette variational (EPV) 
ansatz.\cite{Mezzacapo:2012}  This EPV study uses a single very broad class of 
entangled-plaquette states as trial wave functions, and finds in this very unbiased way that
along the $J_{3}=0$ line the model has N, QP, and aN phases with critical points very close to ours.
In the same region studied by us (viz., $J_{2}<1$) the EPV study finds no evidence of spiral order
along the $J_{3}=0$ line.

It is clear that the spin-$\frac{1}{2}$ $J_{1}$--$J_{2}$--$J_{3}$ HAFM on the honeycomb 
lattice is a challenging model, but one in which there seems now to be a growing
consensus on its overall phase structure.  There is very good agreement over
the regions in which the N and S phases exist, and we believe our own
CCM results now give perhaps the best quantitative results in these cases for
the positions of the phase boundaries and the positions of the two tricritical QCPs 
at $(J_{2}^{c_1},J_{3}^{c_1})$ and $(J_{2}^{c_2},J_{3}^{c_2})$.  

An uncertainty remains over the precise
extent of the phase with spiral order, and whether or not there is an aN phase along the 
$J_{3}=0$ line for values of $J_2$ beyond the point where the QP phase disappears, and hence
also for small positive values of $J_3$ up to the point where spiral order sets in.
The other main uncertainty is the nature of the QP phase itself.  We have argued here
that over at least some widely separated points on the boundaries with the N, S, and
aN phases the QP phase has PVBC order.  Two quite separate ED 
calculations\cite{Albuquerque:2011,Mosadeq:2011} also give clear evidence that much
of the QP phase has PVBC order, although the latter calculations\cite{Mosadeq:2011} are
only done along the $J_{3}=0$ line.  

By contrast the EPV calculations\cite{Mezzacapo:2012} 
along the $J_{3}=0$ line seem to favor a disordered (spin-liquid) phase, while spin-wave
calculations\cite{Mulder:2010} favor SDVBC order along the same line in the QP regime.
The PFFRG study,\cite{Reuther:2011} also done over the entire $J_{2},J_{3} \in [0,1]$
window, finds evidence too that a large part of the QP regime has strong SDVBC order,
while the part with smaller values of $J_2$ has only weak PVBC order.  On the other hand the
SE study\cite{Oitmaa:2011} finds that SDVBC order is not favored, at least for low values
of $J_3$.  

We should note, however, that the SE study is in broad disagreement with most other 
studies along the $J_{3}=0$ line, in that it finds no evidence at all for a
magnetically disordered phase there, but instead finds that the N phase first gives way 
to the second classical spiral phase, and then later to the spiral phase considered here,
as $J_2$ is increased.  On the other hand, the SE results are consistent with the
finding from the ED+SCCMF analysis\cite{Albuquerque:2011} that at least for some parameter ranges the SDVBC
state might be very difficult to distinguish from magnetically ordered states such 
as our S state.

Finally we note that the ED+SCCMF study also presents evidence for the N to PVBC
transition being a strong candidate for a deconfined transition, just as we have found
in our earlier CCM studies of the model along the $J_{3}=J_{2}$ line\cite{DJJF:2011_honeycomb} 
and along the $J_{3}=0$ line.\cite{PHYLi:2012_honeyJ1-J2}  Clearly this model still
has open questions, but we believe that the CCM results presented here have furthered our
understanding of it.

\section*{ACKNOWLEDGMENTS}
We thank J.~Richter for fruitful discussions.  
We are also grateful to
J. Schulenburg for his assistance in
the updating and maintenance of the CCM computer code.  We thank the
University of Minnesota Supercomputing Institute for the grant of
supercomputing facilities.


\begin{thebibliography}{200}

\bibitem{Ramirez:2008}
A.~P.~Ramirez,
Nat.\ Phys.\ {\bf 4}, 442 (2008).

\bibitem{Balents:2010}
L.~Balents, 
Nature (London) {\bf 464}, 199 (2010).

\bibitem{Anderson:1973}
P.~W.~Anderson,  
Mater.\ Res.\ Soc.\ Bull.\ {\bf 8}, 153 (1973).

\bibitem{Anderson:1987}
P.~W.~Anderson,  
Science {\bf 235}, 1196 (1987).

\bibitem{shastry1} 
B.~S.~Shastry and B.~Sutherland, 
Physica B {\bf 108}, 1069 (1981).

\bibitem{Fazekas:1974}
P.~Fazekas and P.~W.~Anderson,  
Phil.\ Mag.\ {\bf 30}, 423 (1974).

\bibitem{Rastelli:1979}
E.~Rastelli, A.~Tassi, and L.~Reatto,
Physica B \& C {\bf 97}, 1 (1979).

\bibitem{Mattson:1994}  
A.~Mattsson, P.~Fr\"ojdh, and T.~Einarsson,
Phys.\ Rev.\ B {\bf 49}, 3997 (1994). 

\bibitem{Fouet:2001}
J.~B.~Fouet, P.~Sindzingre, and C.~Lhuillier, 
Eur.\ Phys.\ J.\ B {\bf 20}, 241 (2001). % full j1-j2-j3 ED

\bibitem{Mulder:2010}
A.~Mulder, R.~Ganesh, L.~Capriotti, and A.~Paramekanti,
Phys.\ Rev.\ B {\bf 81}, 214419 (2010).

\bibitem{Cabra:2011}
D.~C.~Cabra, C.~A.~Lamas, and H.~D.~Rosales,
Phys.\ Rev.\ B {\bf 83}, 094506 (2011).

\bibitem{Ganesh:2011}
R.~Ganesh, D.~N.~Sheng, Y.-J.~Kim, and A.~Paramekanti,
Phys.\ Rev.\ B {\bf 83}, 144414 (2011).

\bibitem{Clark:2011}
B.~K.~Clark, D.~A.~Abanin, and S.~L.~Sondhi,
Phys.\ Rev.\ Lett.\ {\bf 107}, 087204 (2011).

\bibitem{Reuther:2011}
J.~Reuther, D.~A.~Abanin, and R.~Thomale, 
Phys.\ Rev.\ B {\bf 84}, 014417 (2011).

\bibitem{DJJF:2011_honeycomb}
D.~J.~J.~Farnell, R.~F.~Bishop, P.~H.~Y.~Li, J.~Richter, and C.~E.~Campbell,
Phys.\ Rev.\ B {\bf 84}, 012403 (2011).

\bibitem{Albuquerque:2011}
A.~F.~Albuquerque, D.~Schwandt, B.~Het\'{e}nyi, S.~Capponi, M.~Mambrini, A.~M.~L\"auchli,
Phys.\ Rev.\ B {\bf 84}, 024406 (2011).   

\bibitem{Mosadeq:2011}
H.~Mosadeq, F.~Shahbazi, and S.~A.~Jafari,
J.\ Phys.: Condens.\ Matter {\bf 23}, 226006 (2011).

\bibitem{Oitmaa:2011}
J.~Oitmaa and R.~R.~P.~Singh,
Phys.\ Rev.\ B {\bf 84}, 094424 (2011).   

\bibitem{Mezzacapo:2012} 
F.~Mezzacapo and M.~Boninsegni, 
Phys. Rev. B {\bf 85}, 060402(R) (2012).

\bibitem{PHYLi:2012_honeycomb_J1neg}
P.~H.~Y.~Li, R.~F.~Bishop, D.~J.~J.~Farnell, J.~Richter, and C.~E.~Campbell,
Phys.\ Rev.\ B {\bf 85}, 085115 (2012).

\bibitem{PHYLi:2012_honeyJ1-J2}
P.~H.~Y.~Li, R.~F.~Bishop, D.~J.~J.~Farnell, and C.~E.~Campbell,
arXiv:1201.3512v1 [cond-mat.str-el] (2012).

\bibitem{PHYLi:2012_Honeycomb_J2neg}
P.~H.~Y.~Li and R.~F.~Bishop, 
arXiv: 1202.6249v1 [cond-mat.str-el] (2012).

\bibitem{kitaev} 
A.~Kitaev, 
Ann.\ Phys.\ (N.Y.) {\bf 321}, 2 (2006);
G.~Baskaran, S.~Mandal, and R.~Shankar, 
Phys.\ Rev.\ Lett.\ {\bf 98}, 247201 (2007);
J.~Chaloupka, G.~Jackeli, and G.~Khaliullin,
{\it ibid.} {\bf 105}, 027204 (2010).

\bibitem{graphene} 
A.~H.~Castro Neto, F.~Guinea, N.~M.~R.~Peres, K.~S.~Novoselov, and A.~K.~Geim,
Rev.\ Mod.\ Phys.\ {\bf 81}, 109 (2009).

\bibitem{meng}
Z.~Y.~Meng, T.~C.~Lang, S.~Wessel, F.~F.~Assaad, and A.~Muramatsu, 
Nature (London) {\bf 464}, 847 (2010).

\bibitem{Yang:2011_hcomb}
H.~Y.~Yang and K.~P.~Schmidt,
Europhys.\ Lett.\ {\bf 94}, 17004 (2011).

\bibitem{Vaezi:2010}
A.~Vaezi and X. G.~Wen,
arXiv:1010.5744v1 [cond-mat.str-el] (2010).

\bibitem{Vaezi:2011}
A.~Vaezi, M.~Mashkoori, and M.~Hosseini,
arXiv:1110.0116v2 [cond-mat.str-el] (2011).

\bibitem{exp} 
S.~Okubo, F.~Elmasry, W.~Zhang, M.~Fujisawa,
T.~Sakurai, H.~Ohta, M.~Azuma, O.~A.~Sumirnova, and N.~Kumada,
J.\ Phys.: Conf.\ Series {\bf 200}, 022042 (2010).

\bibitem{Miura:2006}
Y.~Miura, R.~Hiari, Y.~Kobayashi, and M.~Sato,
J.\ Phys.\ Soc.\ Jpn.\ {\bf 75}, 084707 (2006).

\bibitem{Kataev:2005}
V.~Kataev, A.~M\"oller, U.~L\"{o}w, W.~Jung, N.~Schittner, M.~Kriener, and A.~Freimuth,
J.\ Magn.\ Magn.\ Mater.\ {\bf 290/291}, 310 (2005).

\bibitem{Regnault:1990}
L.~P.~Regnault and J.~Rossat-Mignod,
in {\it Phase Transitions in Quasi-Two-Dimensional Planar Magnets},
edited by L.~J.~De Jongh (Kluwer Academic Publishers, Dordrecht, 1990), p.~271.

\bibitem{Tsirlin:2010}
A.~A.~Tsirlin, O.~Janson, and H.~Rosner,
Phys.\ Rev.\ B {\bf 82}, 144416 (2010).

\bibitem{Struck:2011}
J.~Struck, C.~{\"O}lsch{\"a}ger, R.~Le Targat, P.~Soltan-Panahi, A.~Eckardt, 
M.~Lewenstein, P.~Windpassinger, and K.~Sengstock,
Science {\bf 333}, 996 (2011).

\bibitem{peps}
V.~Murg, F.~Verstraete, and J.~I.~Cirac,
Phys.\ Rev.\ B {\bf 79}, 195119 (2009).

\bibitem{schulz}
H.~J.~Schulz, T.~A.~L.~Ziman, and D.~Poilblanc,
J.\ Phys.\ I {\bf 6}, 675 (1996).

\bibitem{Richter:2010_ED}
J.~Richter and J.~Schulenberg,
Eur.\ Phys.\ J.\ B {\bf 73}, 117 (2010).

\bibitem{Reuther:2011_J1J2J3mod}
J.~Reuther, P.~W\"{o}lfle, R.~Darradi, W.~Brenig, M.~Arlego, and J.~Richter,
Phys.\ Rev.\ B {\bf 83}, 064416 (2011).

\bibitem{Kr:2000} 
S.~E.~Kr\"uger, J.~Richter, J.~Schulenburg, 
D.~J.~J. Farnell, and R.~F. Bishop,
Phys.\ Rev.\ B {\bf 61}, 14607 (2000).

\bibitem{rachid05}
R.~Darradi, J.~Richter, and D.~J.~J.~Farnell,
Phys.\ Rev.\ B {\bf 72}, 104425 (2005).
% Shastry Sutherland

\bibitem{schmalfuss}  
D.~Schmalfu{\ss}, R.~Darradi, J.~Richter, J.~Schulenburg, and D.~Ihle,
Phys.\ Rev.\ Lett.\ {\bf 97}, 157201 (2006).

\bibitem{Bi:2008_PRB} 
R.~F.~Bishop, P.~H.~Y.~Li, R.~Darradi, J.~Schulenburg, and J.~Richter,
Phys.\ Rev.\ B {\bf 78}, 054412 (2008).

\bibitem{Bi:2008_JPCM} 
R.~F.~Bishop, P.~H.~Y.~Li, R.~Darradi, and J.~Richter,   
J.\ Phys.: Condens.\ Matter {\bf 20}, 255251 (2008).

\bibitem{darradi08} 
R.~Darradi, O.~Derzhko, R.~Zinke, J.~Schulenburg, S.~E.~Kr\"uger, and J.~Richter,
Phys.\ Rev.\ B {\bf 78}, 214415 (2008).

\bibitem{Bishop:2009}
R.~F.~Bishop, P.~H.~Y.~Li, D.~J.~J. Farnell, and C.~E.~Campbell,
Phys.\ Rev.\ B {\bf 79}, 174405 (2009).

\bibitem{richter10}  
J.~Richter, R.~Darradi, J.~Schulenburg, D.~J.~J. Farnell, and H.~Rosner,
Phys.\ Rev.\ B {\bf 81}, 174429 (2010).

\bibitem{UJack_ccm}
R.~F.~Bishop, P.~H.~Y.~Li, D.~J.~J.~Farnell, and C.~E.~Campbell,
Phys.\ Rev.\ B {\bf 82}, 024416 (2010).

\bibitem{Bishop:2012_checkerboard}
R.~F.~Bishop, P.~H.~Y.~Li, D.~J.~J.~Farnell, J.~Richter, and C.~E.~Campbell,
arXiv:1202.2722v1 [cond-mat.str-el] (2012).

\bibitem{ccm2}
C. Zeng, D.~J.~J. Farnell, and R.~F.~Bishop, 
J.\ Stat.\ Phys.\ {\bf 90}, 327 (1998).

\bibitem{j1j2_square_ccm1} 
R.~F.~Bishop, D.~J.~J.~Farnell, and J.~B.~Parkinson, 
Phys.\ Rev.\ B {\bf 58},  6394 (1998).

\bibitem{ccm3} 
R.~F.~Bishop, D.~J.~J.~Farnell, S.~E.~Kr\"uger, J.~B.~Parkinson, J.~Richter, and C.~Zeng, 
J.\ Phys.: Condens.\ Matter {\bf 12}, 6887 (2000).

\bibitem{ccm} 
We use the program package CCCM of D.~J.~J. Farnell and J. Schulenburg, see
http://www-e.uni-magdeburg.de/jschulen/ccm/index.html.


\end{thebibliography}
\end{document}